%% file: main.tex
\documentclass[a4paper,11pt]{article}
\pdfoutput=1 

\usepackage{jinstpub} 
\usepackage[utf8]{inputenc}
\usepackage{lineno}

\usepackage[capitalise,noabbrev,nameinlink]{cleveref}
\newlength{\abc}
\AtBeginDocument{%
\renewcommand{\ref}[1]{\mbox{\Cref{#1}}}

}

\input{setup/packages.tex}
\input{setup/commands.tex}

\title{\boldmath Neutron Irradiation and Electrical Characterisation of the First 8'' Silicon Pad Sensor Prototypes for the CMS Calorimeter Endcap Upgrade}

\author{CMS HGCAL Collaboration}

\emailAdd{thorben.quast@cern.ch}

\abstract{
    As part of its HL-LHC upgrade program, the CMS collaboration is replacing its existing endcap calorimeters with a high-granularity calorimeter (CE). 
    The new calorimeter is a sampling calorimeter with unprecedented transverse and longitudinal readout for both electromagnetic and hadronic compartments. 
    Due to its compactness, intrinsic time resolution, and radiation hardness, silicon has been chosen as active material for the regions exposed to higher radiation levels. 
    The silicon sensors are fabricated as \SI{20}{\centi\metre} (8'') wide hexagonal wafers and are segmented into several hundred pads which are read out individually. 
    As part of the sensor qualification strategy, 8'' sensor irradiation with neutrons has been conducted at the Rhode Island Nuclear Science Center (RINSC) and followed by their electrical characterisation in 2020-21. 
    The completion of this important milestone in the CE's R$\&$D program is documented in this paper and it provides detailed account of the associated infrastructure and procedures.
    The results on the electrical properties of the irradiated CE silicon sensors are presented.
} 

\begin{document}
\maketitle
\flushbottom

\input{content/1_introduction.tex}
\input{content/2_sensors.tex}
\input{content/2_irradiation.tex}
\input{content/4_setup.tex}
\input{content/5_results.tex}
\input{content/7_conclusion.tex}
\input{content/acknowledgments.tex}

\appendix
\input{content/appendix/irradiation_rounds.tex}

\input{content/appendix/chuck_temperature.tex}

\bibliographystyle{unsrt}
\bibliography{bib/bib}

\end{document}

%% file: setup/packages.tex
\usepackage[binary-units=true]{siunitx}

\usepackage{multirow}

\usepackage{textcomp}
\usepackage{xcolor}
\usepackage{graphicx}
\usepackage{mathtools}
\usepackage{subcaption}
\usepackage[LGRgreek]{mathastext}
\usepackage{xspace}

%% file: setup/commands.tex
\newcommand{\ind}[0]{\hspace*{\customindent}}

\newcommand{\parag}{\parag\ind}

\newcommand{\tocite}[1]{\textbf{\textcolor{red}{[\ifthenelse{\equal{#1}{}}{?}{#1}]}}}

\newcommand{\percent}{$\%$}

\newcommand{\neqcm}{\ensuremath{\mathrm{n_{eq}}/\mathrm{cm}^{2}}\xspace}

\newcommand{\micron}{\ensuremath{\upmu\mathrm{m}}\xspace}

\definecolor{ttHcol}{RGB}{67,118,201}
\definecolor{ttbbcol}{RGB}{235,230,10}
\definecolor{ttbcol}{RGB}{205,0,10}
\definecolor{tttbcol}{RGB}{255,153,1}
\definecolor{ttcccol}{RGB}{131,38,10}
\definecolor{ttlfcol}{RGB}{81,142,25}
\definecolor{Singletcol}{RGB}{0,204,204}
\definecolor{Vjetscol}{RGB}{104,140,140}
\definecolor{Dibosoncol}{RGB}{1,25,147}
\definecolor{ttVcol}{RGB}{255,102,101}

%% file: content/1_introduction.tex
\section{Introduction}
\label{sec:introduction}
In the current decade, the Large Hadron Collider~\cite{evans:2008} (LHC) at CERN will be upgraded to the High-Luminosity LHC (HL-LHC)~\cite{hl-lhc-tdr:2017}.
Its instantaneous luminosity is designed to reach at least five times the LHC's design value.
The increase in both the expected number of pile-up interactions and the anticipated damage due to the enhanced radiation levels will pose significant challenges to the LHC experiments.\newline
The CMS~\cite{cms:2008} collaboration has undertaken an extensive R$\&$D program to upgrade many parts of the detector.
One of the planned upgrades consists of the replacement of the existing endcap calorimeters with a high-granularity calorimeter (CE)~\cite{hgcal-tdr:2018}.
The new CE will consist of 47 sampling layers interspersed with absorber plates and will feature close to six million readout channels.
For its compactness, fast signal formation, and acceptable radiation hardness, silicon has been chosen as one of the active materials.
Approximately $\SI{600}{\metre\squared}$ of it will be deployed in the regions where integrated fluences up to $1 \times 10^{16}~$1-MeV neutron-equivalents per square centimeter ($\neqcm$) 
at the end of its 10-year operation at the HL-LHC are expected. 
Prototypes of silicon-based modules have already been built and were tested with particle beams~\cite{cms_hgc-2016-beamtests,H1:2020,H2:2020,H3:2021}.\newline
The silicon sensors will be fabricated as multi-pad, DC-coupled, n-on-p ("p-type"), 8'' hexagonal wafers with active thicknesses of \SI{120}{\micro\metre}, \SI{200}{\micro\metre} or \SI{300}{\micro\metre}.
To assess their anticipated degradation over the HL-LHC lifetime, irradiation studies are an integral component to the CE's R$\&$D efforts.
However, the irradiation and electrical characterisation of full 8'' silicon sensors are not trivial due to their large size and the large number of pads.
In fact, previous R$\&$D on the CE silicon sensors' bulk degradation in terms of their leakage currents and depletion voltages was, thus far, limited to testing of small, approximately \SI{1}{\centi\metre\squared}-sized, single-diode test structures~\cite{Curr_s_2017,Akchurin:2020}.
As part of the CE's prototyping phase in the last few years, the CMS collaboration has developed the required infrastructure and procedures for the irradiation and the electrical characterisation of its 8'' silicon sensors.
The neutron irradiation was performed at the Rhode Island Nuclear Science Center (RINSC) which offers the necessary infrastructure to expose such large objects up to  $\sim 10^{16}~\neqcm$ within a few hours.
Afterwards, the ARRAY system~\cite{pitters:array2019} was used for the electrical characterisation of all $\mathcal{O}(100)$ pads on the silicon sensors.
In order to reduce the fluence-induced leakage current to measurable levels, the tests were conducted at \SI{-40}{\celsius}.\newline
This paper is structured as follows:
\ref{sec:sensors} deals with the CE silicon pad sensor prototypes in terms of their design and the production parameters.
\ref{sec:irradiation} describes the neutron irradiation of the 8'' wafers at RINSC. 
The electrical test setup based on the ARRAY system and the measurement procedure are documented in~\ref{sec:setup}.
Results on the electrical sensor characteristics and their limitations are discussed in~\ref{sec:results}.
Finally, the conclusions are given in~\ref{sec:conclusion}.

%% file: content/2_sensors.tex
\section{Silicon Pad Sensor Prototypes for the CMS Endcap Calorimeter Upgrade}
\label{sec:sensors}
\begin{figure}
	\captionsetup[subfigure]{aboveskip=-1pt,belowskip=-1pt}
	\centering
	\begin{subfigure}[b]{0.50\textwidth}
		\includegraphics[width=0.999\textwidth]{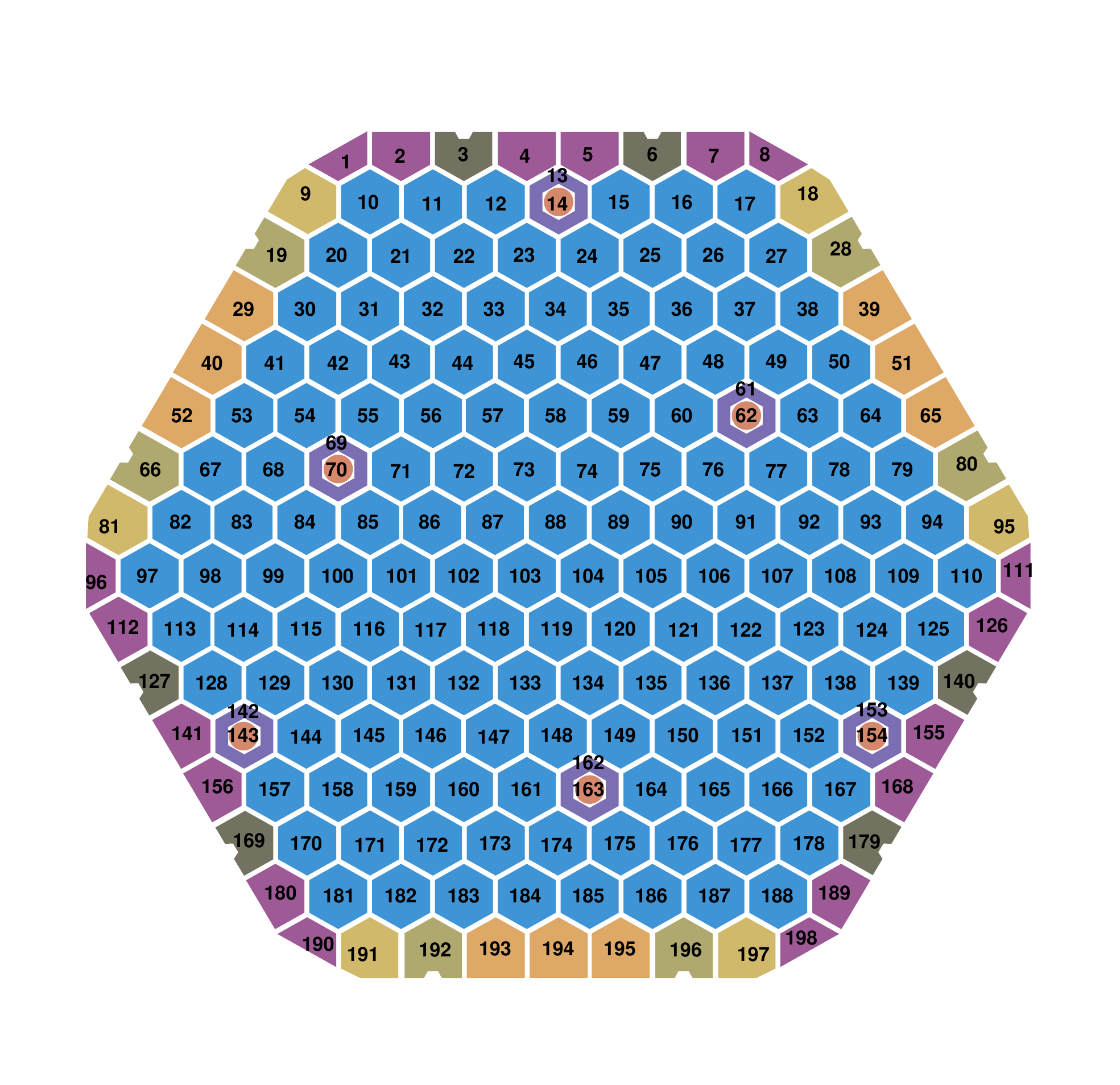}
		\subcaption{
		}
	\end{subfigure}
	\hfill
	\begin{subfigure}[b]{0.48\textwidth}
		\includegraphics[width=0.999\textwidth]{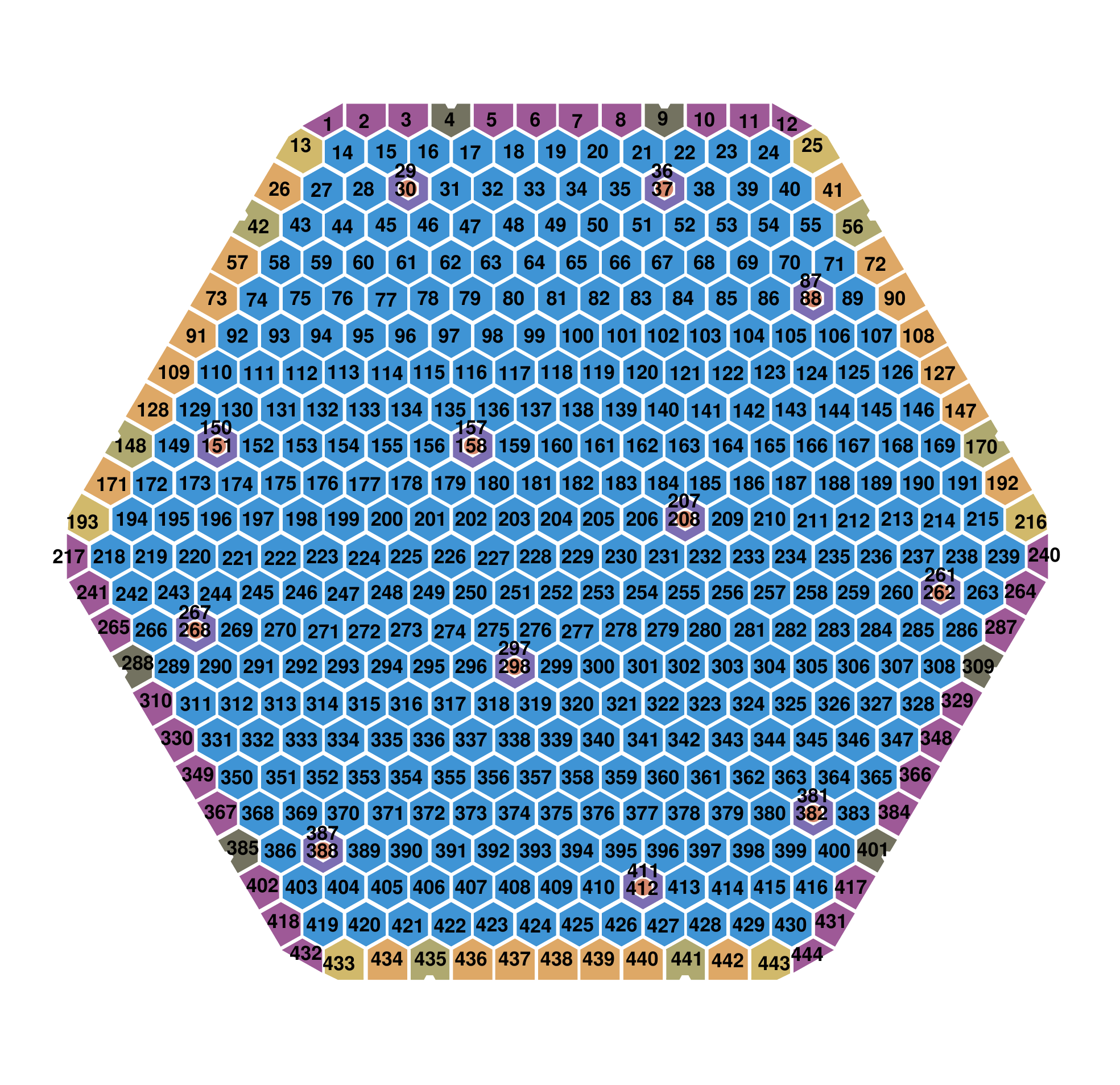}
		\subcaption{
		}
	\end{subfigure}    
	\caption{
        Layout of the tested 8'' prototype silicon pad sensors with (a) the low-density (LD) and (b) the high-density (HD) design.
		Regular-sized hexagonal pads are depicted in cyan.
		Edge and corner pads populate the sensor periphery. 
		Six (LD), respectively twelve (HD), regular, small hexagonal pads on the sensor are dedicated for energy calibration with minimum ionising particles up to the end of life.
		The colour coding highlights the usage of different pad geometries.
		The sensor guard rings are not shown.
	}
	\label{fig:Sensors}
\end{figure}

The CMS high granularity calorimeter will be made of more than \SI{600}{\metre\squared} of planar DC-coupled silicon pad sensors.
Starting from 8'' circular wafers, the sensors are diced into hexagons.
P-type doping of their bulk was chosen due to empirical evidence of superior noise performance after irradiation with respect to n-type doping~\cite{Adam_2017}.
The oxygen content of those 8” wafers is reduced with respect to the case of the 6” wafers used for the CMS and ATLAS trackers which, in principle, could impair the sensor's radiation hardness.\newline
In this work, the first version prototypes of hexagonal wafer sensors with the so-called low-density (LD) and high-density (HD) designs were irradiated with neutrons and electrically characterised.
Their design is illustrated in~\ref{fig:Sensors}.
Each CE silicon sensor is segmented into several hundred pads that constitute the sensitive units. 
The majority of those pads are shaped as regular hexagons which are drawn as cyan-colored pads in~\ref{fig:Sensors}.
Special non-hexagonal structures fill out the sensor periphery.
The arrangement of pads on a sensor is enclosed by a set of three guard rings (grounded, floating, biased) which protect the sensor from currents drawn from its dicing edges.
The LD sensors were produced from physically thinned p-type float zone silicon wafers.
They are segmented into 198 pads, where full hexagonal pads are about \SI{1.2}{\centi\metre\squared}, and \SI{200}{\micro\meter} or \SI{300}{\micro\meter} in active thickness.
LD sensors will be installed in regions of intermediate radiation levels inside the CE.
By contrast, regions with the highest radiation levels will be populated with HD sensors whose active thickness is \SI{120}{\micro\meter}.
HD sensors are segmented into 444 pads (full hexagon pad size: \SI{0.5}{\centi\metre\squared}) and are produced by epitaxial growth.\newline
As it is foreseen for the final design, p-stop structures were added to all the tested prototype silicon sensors in this work in order to limit the accumulation of electrons on the Si/SiO$_2$ interface after exposure to high radiation levels which otherwise could form a conduction channel, effectively shorting adjacent pads.
Those structures were either bound to the single pads (individual p-stop) or shared between neighboring ones (common p-stop), representing two options under consideration for the CE.
In addition, the tested prototype sensors differed in their flatband voltage (pre-irradiation value of either \SI{-2}{\volt} or \SI{-5}{\volt}).
Unless explicitly stated otherwise, all other production parameters, such as doping concentrations and the production process for the oxide material, were identical for all prototype sensors discussed in this paper.\newline
Prior to irradiation, the sensors had been electrically qualified and proper functionality could be verified.
In particular, full depletion was achieved around \SI{40}{\volt} for \SI{120}{\micro\metre} sensors, \SI{120}{\volt} for \SI{200}{\micro\metre} sensors, and \SI{280}{\volt} for \SI{300}{\micro\meter} sensors. 
Moreover, per-pad leakage currents of the non-irradiated sensors did not exceed a few nA, and the total currents at room temperatures between 20--\SI{24}{\celsius} over the full sensor remained well below \SI{100}{\micro\ampere} for (absolute) bias voltages up to \SI{850}{\volt}.\newline
In general, exposure of silicon sensors to radiation causes displacement damage to the silicon lattice which affects both the leakage currents and the depletion voltages. 
While the bulk-dominated leakage current density increases proportionally with the fluence, the expected increase of the depletion voltage (for p-type sensors) is non-trivial and its quantification, like it is done e.g. in Refs.~\cite{moll:SiDamages,LINDSTROM200330}, is beyond the scope of this work.

%% file: content/2_irradiation.tex
\section{Irradiation at the Rhode Island Nuclear Science Center}
\label{sec:irradiation}

\subsection{Rhode Island Nuclear Reactor}
\label{subsec:RINSC}
\begin{figure}
  \begin{center}
    \includegraphics[width=0.99\textwidth]{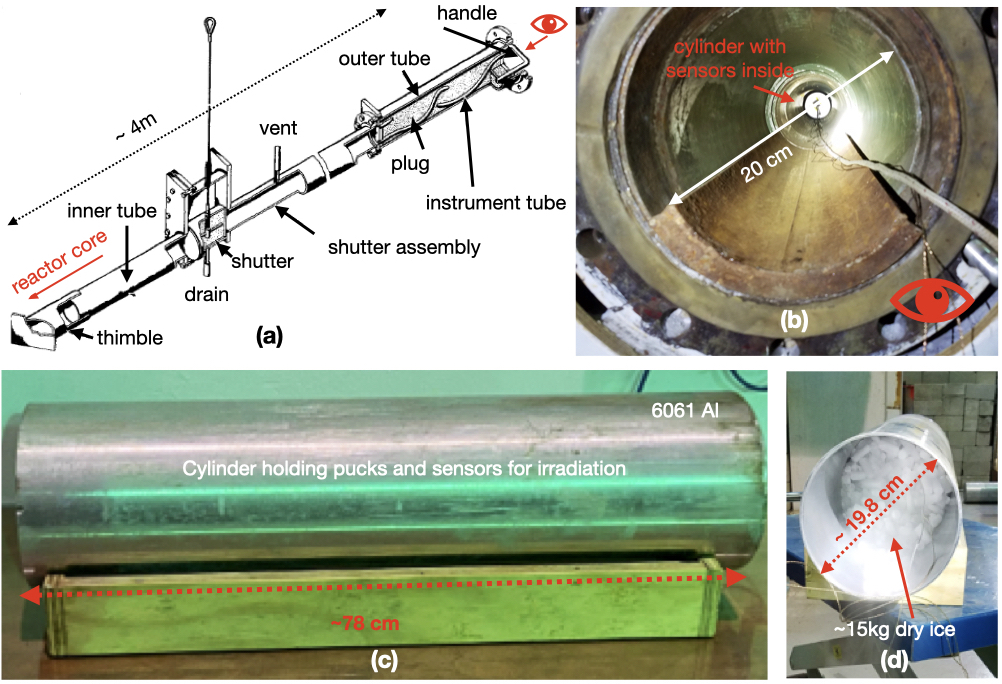}
    \caption{(a) Schematic of the beam port sample delivery system at RINSC.
    (b) The view into the beam port used for these irradiation studies.
    (c) The sample delivery cylinder that contains the sensor-holding hockey pucks and (d) dry ice for cooling of the silicon sensors in the beam port during and after irradiation.
    }
    \label{fig:Beamport_Schematic}
  \end{center}
\end{figure}

The Rhode Island Nuclear Science Center (RINSC) houses a \SI{2}{\mega\watt}, light-water cooled, pool-type reactor in Narragansett, Rhode Island, USA.
Its core consists of fuel assemblies moderated with a combination of graphite and beryllium.
The fuel is plate type U$_3$Si$_2$, cladded with aluminum, enriched to less than 20$~\%$ Uranium-235.\newline
The reactor beam port was used as a sample delivery system because it was the only reactor access point large enough to fit a full-sized CE sensor.
The beam port in question had not previously been used for experiments, only for facility-related work.
A sketch of the beam port and a photo of its inside are shown in \ref{fig:Beamport_Schematic}.
It measures about \SI{4}{\metre} from its opening to the termination near to the reactor core, and it can accommodate samples with diameters of up to \SI{20}{\centi\metre} and with depths up to \SI{90}{\centi\metre}.
A shutter assembly is located \SI{3}{\metre} from the opening of the beam port, which must be raised to allow for the insertion of the samples and closed prior to the start of operations.
A \SI{85}{\centi\metre}-long lead plug serves as a radiation shield that has to be inserted into the opening of the beam port prior to the start of the irradiation.

\subsection{Irradiation of CE Silicon Sensor Prototypes}
\label{subsec:irradiation}
\begin{figure}
  \begin{center}
    \includegraphics[width=0.99\textwidth]{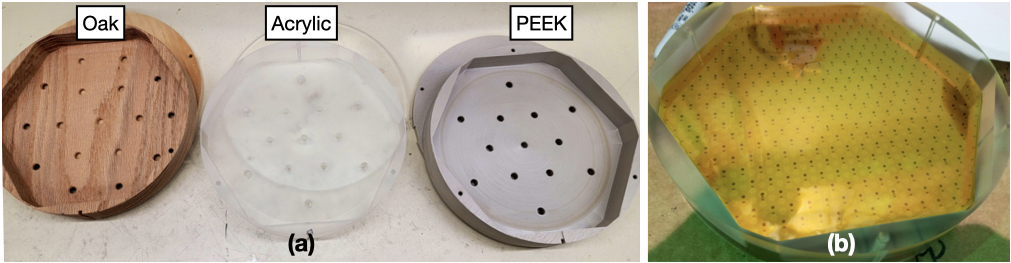}
    \caption{(a) Sample containers ('hockey puck') designed to hold CE silicon sensors for neutron irradiation at RINSC. 
    Wood (oak), acrylic, and PEEK were the puck materials that were experimented with.
    (b) An 8'' high-density CE prototype silicon sensor inside an acrylic puck covered with a Kapton\texttrademark$~$foil.}
    \label{fig:Pucks_Arrayed}
  \end{center}
\end{figure}

\begin{figure}
  \begin{center}
    \includegraphics[width=0.69\textwidth]{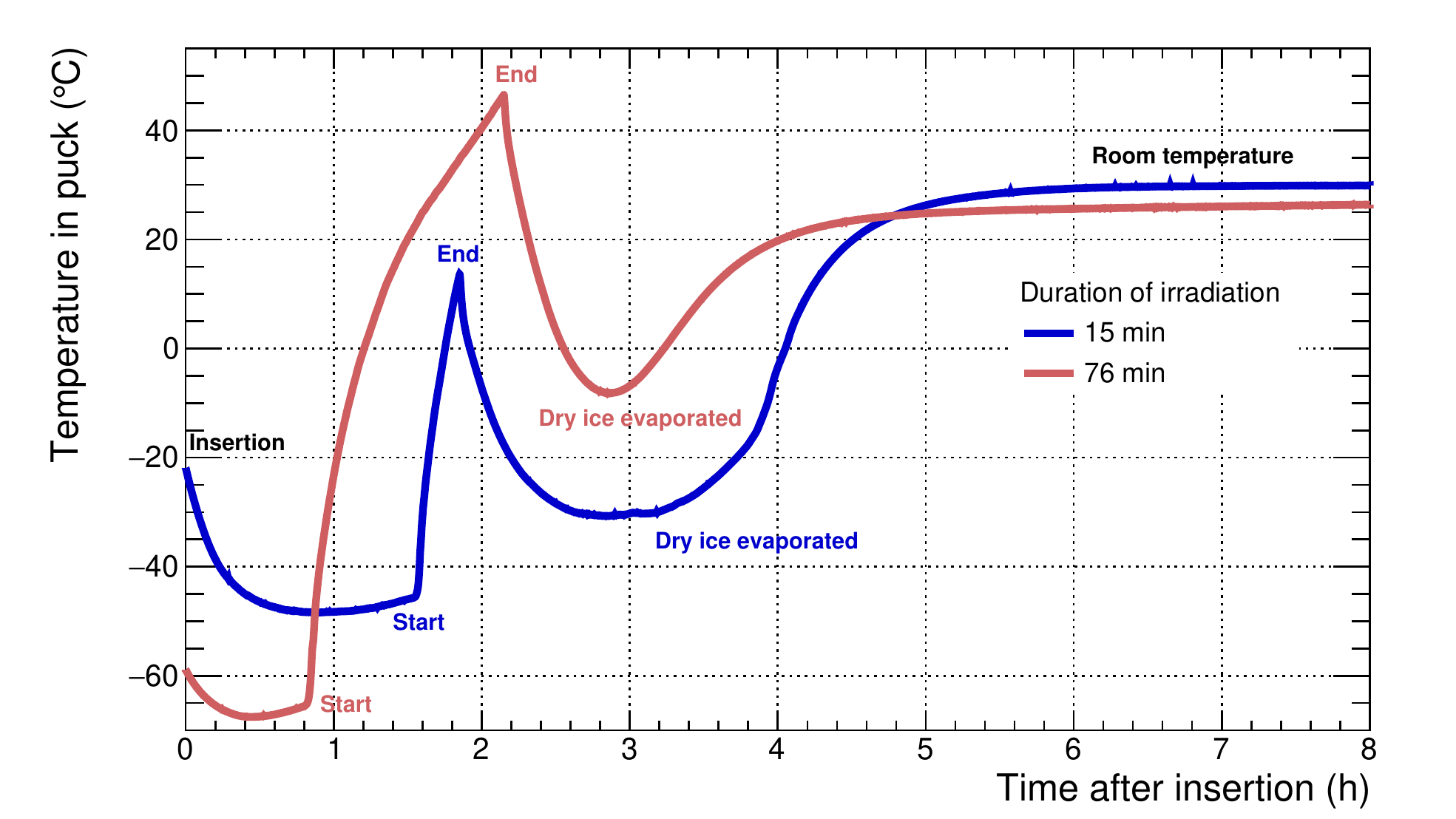}
    \caption{Representative temperature recordings during two of the RINSC irradiation rounds of CE prototype silicon sensors. 
    The time when the irradiation was started and when it ended is indicated.
    The increase of the temperature towards its plateau coincides with the sublimation of the dry ice inside the delivery cylinder.
    }
    \label{fig:Round_10_Temperature_Profile}
  \end{center}
\end{figure}

Given the constraints of the beam port system at RINSC, a dedicated sample delivery method had to be developed for irradiating CE silicon sensors.
This new sample delivery allows for:
\begin{itemize}
  \item positioning sensors as close to the reactor core as possible;
  \item protecting them from physical damage during the loading, irradiation, and unloading;
  \item keeping the irradiated sensors at low temperatures during and after the irradiation;
  \item monitoring the temperature of the samples inside the beam port.
\end{itemize}
Two compatible pieces of hardware were manufactured: a sensor container, referred to as a "hockey puck", and a sample delivery cylinder (see~\ref{fig:Beamport_Schematic}). 
The former is used to protect, orient, and store the sensors during the irradiation while the latter is used to protect and locate the hockey puck inside the beam port.
The cylinder is made from 6061 aluminum and has an outer(inner) diameter of \SI{19.8}{\centi\metre}(\SI{19.1}{\centi\metre}), allowing for a snug fit of the cylinder inside the beam port and a smooth insertion and removal of the hockey pucks.
Three different hockey puck materials were considered (see~\ref{fig:Pucks_Arrayed}): oak, acrylic, and PEEK. 
They differ in their mechanical properties with respect to temperature and humidity fluctuations, in their activation, radiation hardness and overall cost.
As a practical compromise between these factors, acrylic was found most suitable for irradiation up to fluences of $5\times 10^{15}~\neqcm$, and PEEK material suitable for higher fluences. 
The usage of oak was eventually discarded because of its unfavorable tendency to absorb moisture.
The puck base has an outer diameter of \SI{18.6}{\centi\metre} that allows for a smooth fit inside the cylinder. 
The interior of the puck is milled out in the profile of the silicon sensors with an additional clearance of \SI{1}{\milli\metre}. 
With these constraints the thinnest sections of the wall of the puck are slightly over \SI{1}{\milli\metre} thick.\newline
Kapton\texttrademark$~$foils were used to separate sensors in a stack such that no sensors were in direct contact with any metallic surfaces (cf.~\ref{fig:Pucks_Arrayed}).
In addition, antistatic foam was used for covering the top and bottom of the sensor-Kapton\texttrademark$~$stack serving as a cushion  against the walls inside the puck.
After preparation of the puck, the latter was inserted into the delivery cylinder.\newline
In general, the silicon sensors should be kept at low temperatures during the irradiation in order to limit unwanted thermal annealing.
For this purpose, the rest of the delivery cylinder was filled with 15-\SI{18}{\kilo\gram} of dry ice (\ref{fig:Beamport_Schematic}).
In order to monitor the temperature, PT1000 resistance temperature detectors (RTD) were inserted into the puck, at the front and back faces, to record the temperature throughout the irradiation for assessment of the expected sensor annealing during irradiation. 
Representative temperature recordings from two irradiation rounds of different duration are shown in~\ref{fig:Round_10_Temperature_Profile}.
During irradiation, the temperature rose significantly, eventually reaching temperatures ranging from a few \SI{10}{\celsius} to \SI{100}{\celsius}.
In fact, it was found that most of the dry ice sublimated.
Only after shutdown of the reactor, the temperature decreased again.
As a result, the annealing of the silicon sensors inside the beam port was not negligible.
The corresponding annealing time at \SI{60}{\celsius}, cf. Ref.~\cite{moll:SiDamages}, varied between a few minutes to a few hundred minutes for all sensors irradiated at RINSC in 2020-21 up to $\sim 10^{16}~\neqcm$.
After \SI{24}{\hour} in the beam port after irradiation, the cylinder's activation levels had decayed sufficiently, it could be safely extracted, and transferred into a storage freezer.

\subsection{Fluence Assessment}
In addition to the sensors, mechanical packing material, and temperature sensors, each puck contained a number of reference samples for measuring the fluence achieved during an irradiation round. 
Two different objects were found appropriate for measuring the fluences during this campaign: silicon diodes, which were studied for usage in the D0 experiment~\cite{D0diodes}, and ultrapure iron foils. 
The diodes were included inside the puck as close to the sensors as possible by encasing them in small plastic bags and taping them to the inside faces of the puck. 
By contrast, the iron foils were attached to the exterior of the cylinder for ease of removal and gamma spectroscopy measurements. 
The latter were performed at RINSC after irradiation for derivation of the integrated fluence.
In addition, CV and IV measurements of the irradiated diodes were performed at Brown University to assess the depletion voltage, the associated dark current and ultimately the fluence assuming the literature value for the current-related leakage current rate of $\left(3.99\pm 0.03\right)\times 10^{-17}~$A/cm at \SI{20}{\celsius}~\cite{moll:SiDamages}.
The reference silicon diodes are most useful for the lower to medium range of the targeted fluences,  where their depletion voltage was well within the measurement range ($<\SI{1000}{\volt}$).
For higher fluences, full depletion of the silicon diodes could not be reached, and the gamma ray spectra derived from the iron foils are considered more reliable.
\ref{table:irrads} in~\ref{appendix:irrad_rounds} shows the estimated actual fluences from those reference samples.
In order to improve the accuracy of the fluence assessment, future irradiations will include iron foils inside the puck as well as additional silicon test structures.

%% file: content/4_setup.tex
\section{Electrical Characterisation of Neutron-Irradiated Silicon Pad Sensors}
\label{sec:setup}
The experimental setup and the electrical characterisation procedure of neutron-irradiated CE silicon pad sensors are presented in this section.
The following paragraphs describe two similar setups installed at CERN and Texas Tech University, which were used to obtain the results reported in this paper. 
Since the majority of the results are based on the CERN setup, its specifics are elaborated in the following.

\subsection{ARRAY- and Cold-Chuck Based Setup at CERN}
\label{subsec:setup_alps}
At CERN, the S200FA probe station produced by Wentworth Laboratories Ltd. was used for the electrical characterisation of neutron-irradiated silicon pad sensors. 
Its chuck (Systems att, C200-40 model) is temperature-controlled down to \SI{-40}{\celsius}.
\begin{figure}
	\centering
	\includegraphics[width=0.75\textwidth]{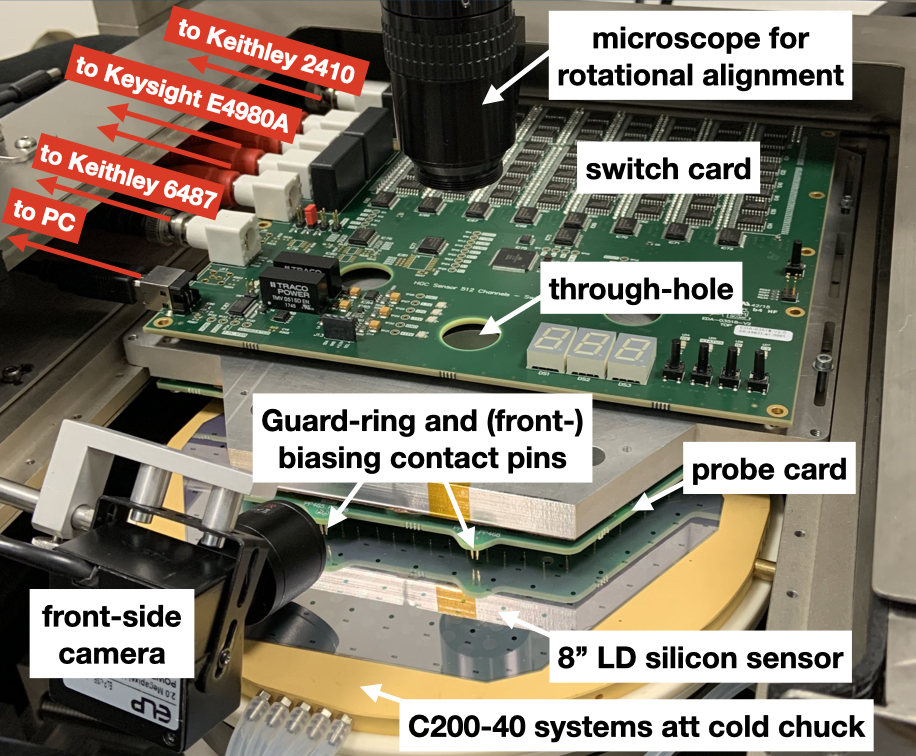}
	\caption{
		A low-density CE silicon pad sensor right before connecting to the switch- and probe-card ARRAY system at CERN.
		The probe station was closed and continuously flushed with dry air during the testing to prevent the formation of ice during low-temperature operation.
		}
	\label{fig:ALPS_setup}
\end{figure}
Apart from a Keithley 2410 power supply, a Keithley 6487 picoammeter, and a Keysight E4980A LCR-meter, all relevant components were installed inside the probe station, see \ref{fig:ALPS_setup}.\newline
The silicon sensors were first placed on the chuck and then connected to the probe- and switch-card based ARRAY system that is described in detail in Ref.~\cite{pitters:array2019}.
The contact between the probe card and the pads was realised with spring-loaded pins.
Through-holes in the cards and the probe station's microscope enable sufficient sensor-to-probe card alignment. 
Also, the sensor's grounded guard ring was accessible via a dedicated contact pad and could be connected with dedicated pins on the probe card.
The high voltage from the power supply was provided to the chuck and with it to the sensor's backside.
Two probe cards specific for the LD and HD sensor layouts have been designed and manufactured.
The switch card was operated with a bias resistance (R$_\text{bias}$) of \SI{1}{\mega\ohm} and a high voltage resistance (R$_\text{HV}$) of \SI{12}{\kilo\ohm}, see$~$\cite{pitters:array2019} (Figure$~$9).
In this configuration, the ARRAY system is designed to safely withstand total leakage currents up to \SI{2}{\milli\ampere} and per-pad currents up to \SI{10}{\micro\ampere}.\newline
Since these limits would have been exceeded by a few orders of magnitude at room temperature, cooling the neutron-irradiated sensors down to \SI{-40}{\celsius} hereby reducing the leakage current by almost three orders of magnitude with respect to room temperature was imperative.
The spatial variation of the C200-40 chuck temperature profile at this temperature amounts to $\pm\SI{0.5}{\celsius}$, cf.~\ref{appendix:chuck_temp}, whereas fluctuations with time were found to be negligible. 
For the purpose of preventing the formation of ice, the probe station was continuously flushed with dry air.\newline 
With total currents at the order of $\mathcal{O}(\SI{1}{\milli\ampere})$, the voltage drop at R$_\text{HV}$ for the testing of neutron-irradiated sensors corresponded to a few volts and was corrected for.
Per-pad leakage currents were measured with the picoammeter, whereas total currents were measured directly with the power supply.
The LCR meter was operated at a frequency (f$_\text{LCR}$) of \SI{2}{\kilo\hertz} for the inference of the per-pad impedance.
This particular frequency was chosen to minimise the error associated to the capacitance that is derived from it~\cite{pitters:array2019}.

\subsection{Characterisation Procedure}
\label{subsec:setup_procedure}
After connecting the sensor to the probecard, per-pad leakage currents as a function of the bias voltage for all pads on a given sensor were measured (IV).
This was followed by a per-pad capacitance \emph{vs.} bias voltage assessment (CV).
After each iteration over all pads at a fixed bias voltage, voltages were incremented in varying steps between 50-\SI{100}{\volt} up to \SI{850}{\volt}, whereby the exact choice depended on the measurement type (IV/CV) and on the thickness of the sensor.
Although not applicable for the results shown in this work, it should be noted that a given measurement sequence was aborted if the total leakage current exceeded \SI{2}{\milli\ampere}.
Similarly, individual pads whose leakage currents exceeded \SI{5}{\micro\ampere} were not measured any further and in particular were excluded from the subsequent CV.
These compliance limits prohibited large voltage drops inside the test circuitry, minimising the risk of damage to the ARRAY system.\newline
The entire characterisation sequence was fully automatised as a LabVIEW-based program (HexDAQ version 1.5.1~\cite{labview_hexdaq}).
Including voltage ramps and settling times, the IV(CV) measurements of low-density sensors in this work took about \SI{1.5}{\hour}(\SI{2.5}{\hour}), whereby the duration is proportional to the number of voltage steps (10-15 here).
The duration was about twice as long for high-density sensors due to the higher number of pads.
In order to quantify the potential performance benefit via annealing, silicon sensors were warmed up to \SI{60}{\celsius} inside the probe station.
After a total of \SI{80}{\min} at this temperature \footnote{Corresponding to about three years at \SI{0}{\celsius}.}, the sensor's depletion voltage is expected to be minimal~\cite{Moll:300958} and the studied sensors were cooled again to \SI{-40}{\celsius}.
IV and CV characterisations were conducted before and after additional annealing.

%% file: content/5_results.tex
\section{Results}
\label{sec:results}
The results of the electrical characterisation of the neutron-irradiated CE silicon pad sensor prototypes are presented in this section. 
\ref{subsec:leakagecurrents} focuses on the discussion of leakage currents (\emph{IV}) whereas \ref{subsec:Udep} addresses the capacitance (\emph{CV}) and depletion voltage assessments. 
\ref{subsec:discussion} offers a discussion on the drawn conclusion.

\subsection{Leakage Current}
\label{subsec:leakagecurrents}
\begin{figure}
	\captionsetup[subfigure]{aboveskip=-1pt,belowskip=-1pt}
	\centering
	\begin{subfigure}[b]{0.49\textwidth}
		\includegraphics[width=0.999\textwidth]{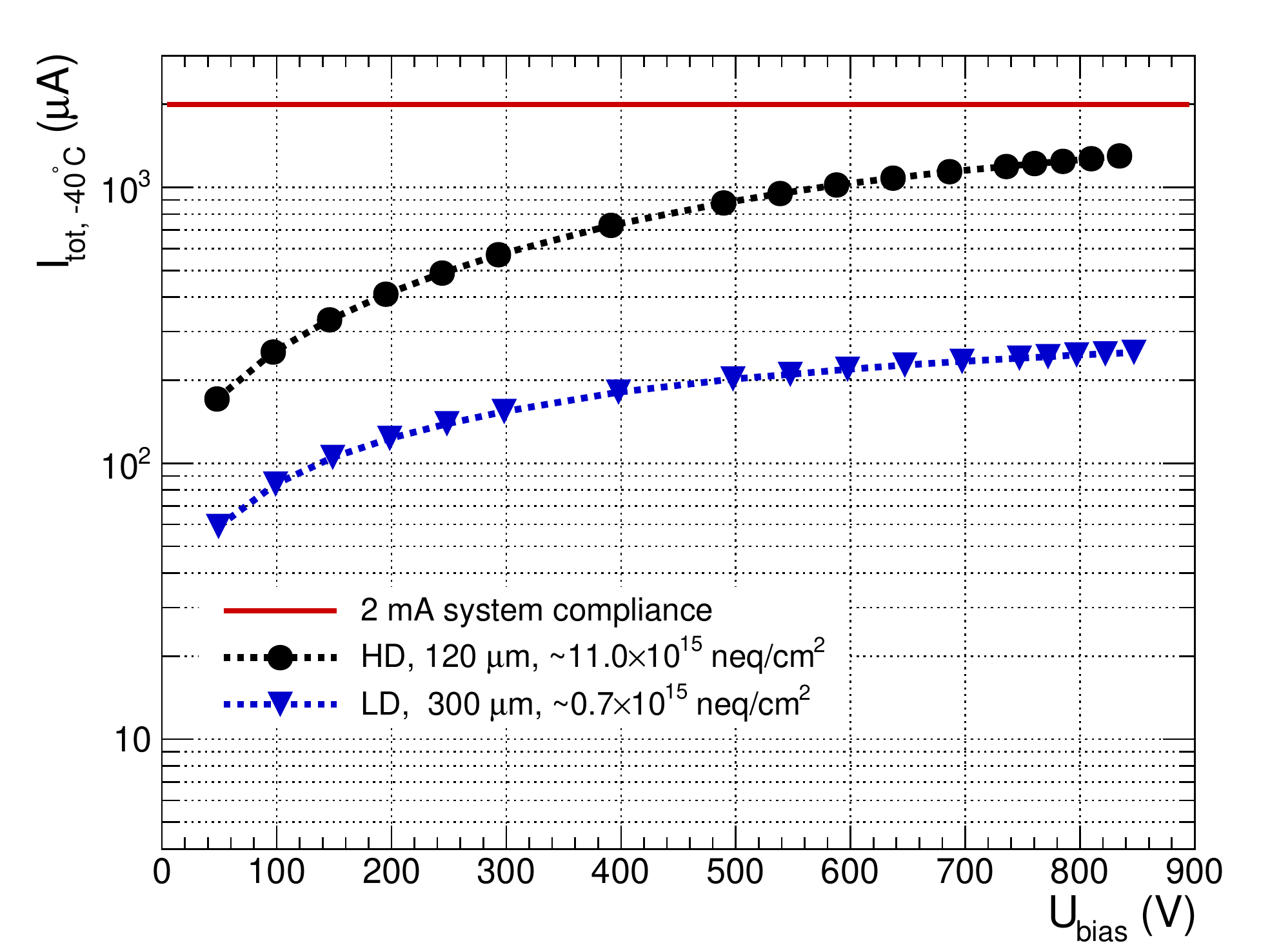}
		\subcaption{
		}
		\label{plot:tot_IV_good}
    \end{subfigure}
    \hfill
    \begin{subfigure}[b]{0.49\textwidth}
        \includegraphics[width=0.999\textwidth]{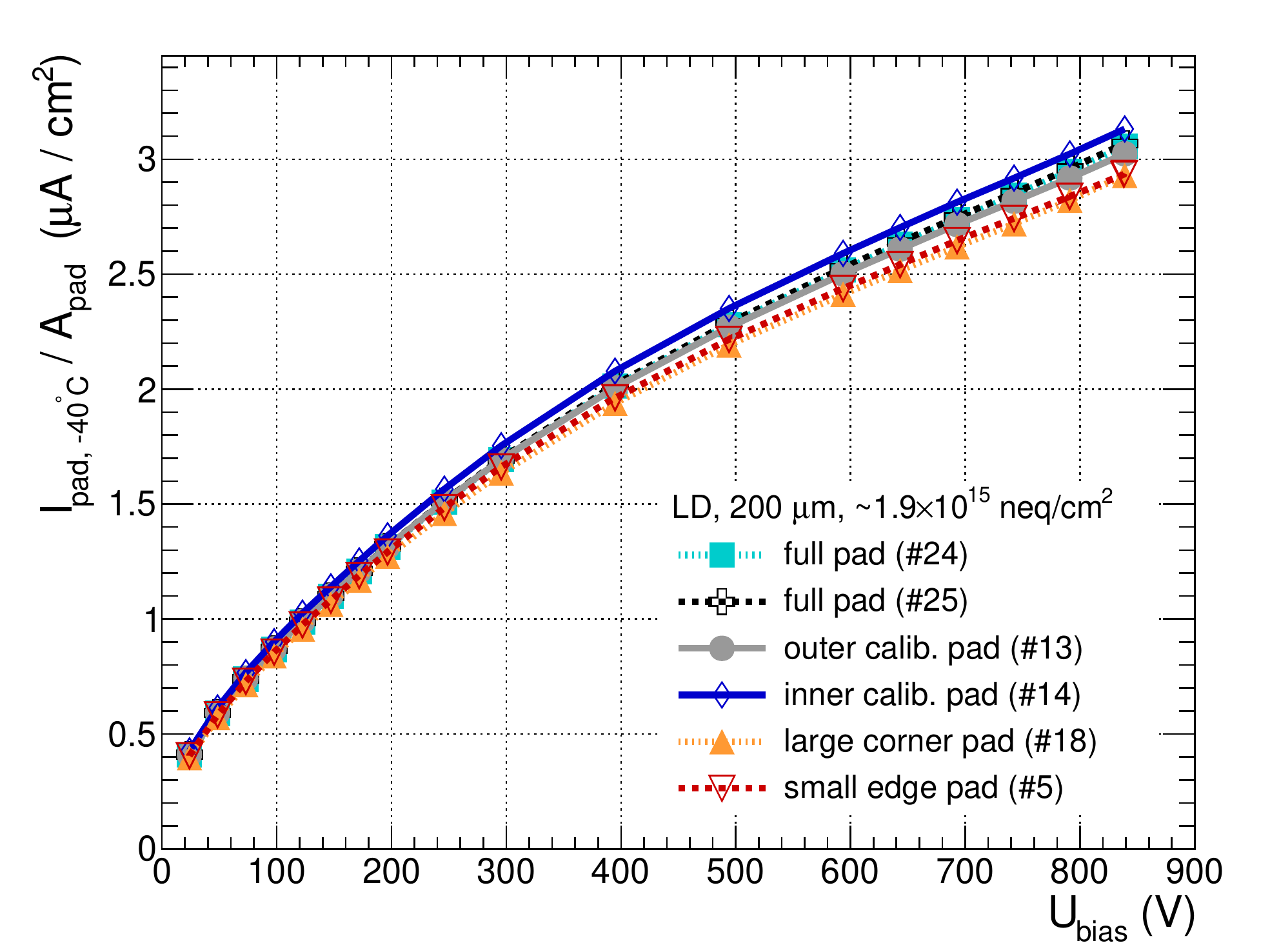}
        \subcaption{
        }
        \label{plot:pad_IV_channels}
    \end{subfigure}

	\caption{
		(a) Total leakage currents after irradiation (without additional thermal annealing) for two representative example sensors. 
		Currents were measured at \SI{-40}{\celsius} ($\text{I}_\text{tot, \SI{-40}{\celsius}}$) and at different effective bias voltages ($\text{U}_\text{bias}$). 
        (b) Per-pad leakage currents normalised to the area of full hexagonal pads as a function of the bias voltage for different pads with different geometries on one example sensor.
	}
\end{figure}
\begin{figure}
	\captionsetup[subfigure]{aboveskip=-1pt,belowskip=-1pt}
	\centering
	\begin{subfigure}[b]{0.49\textwidth}
		\includegraphics[width=0.999\textwidth]{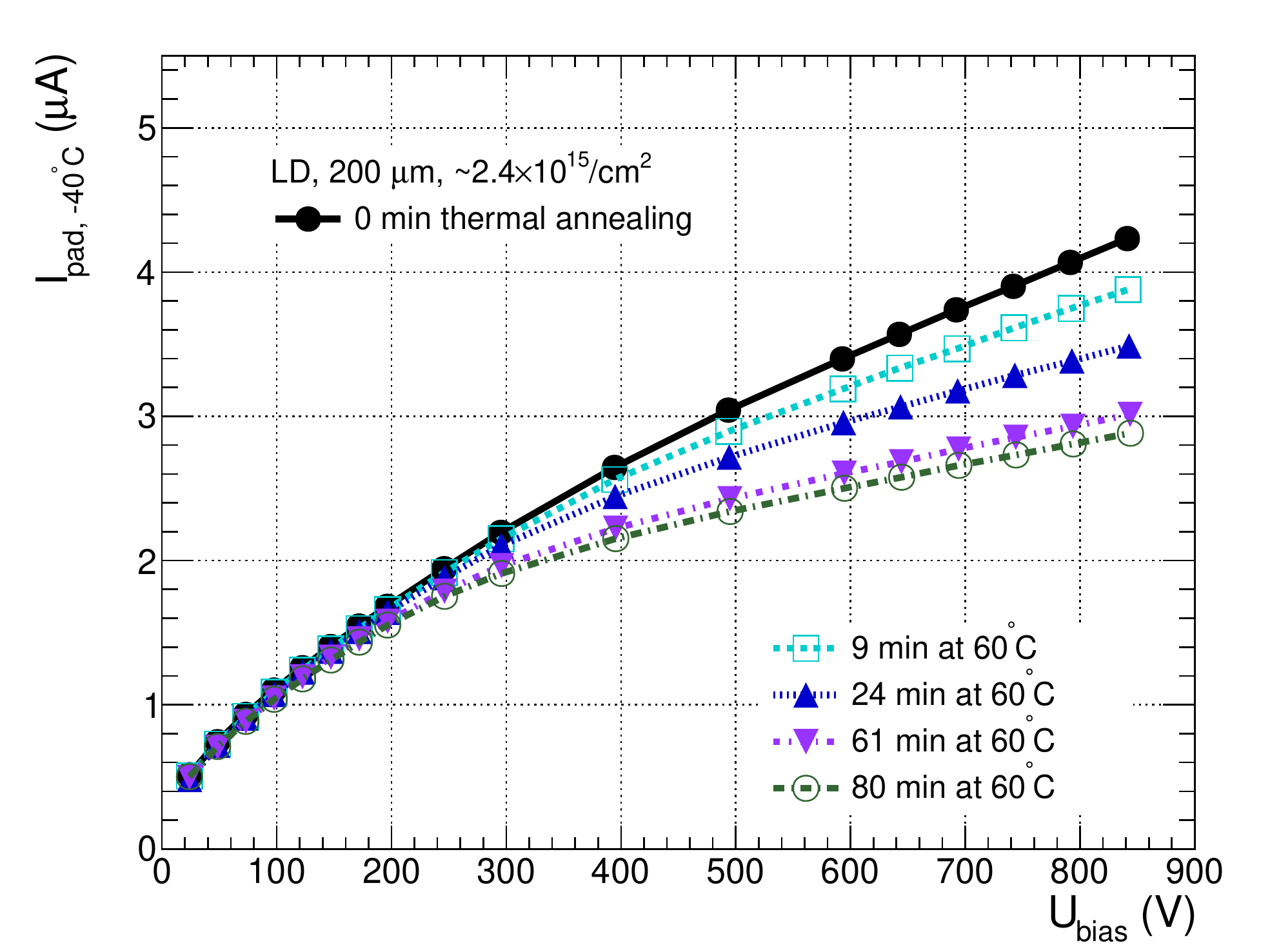}
		\subcaption{
		}
		\label{plot:annealing_IV}
	\end{subfigure}
	\hfill
	\begin{subfigure}[b]{0.49\textwidth}
		\includegraphics[width=0.999\textwidth]{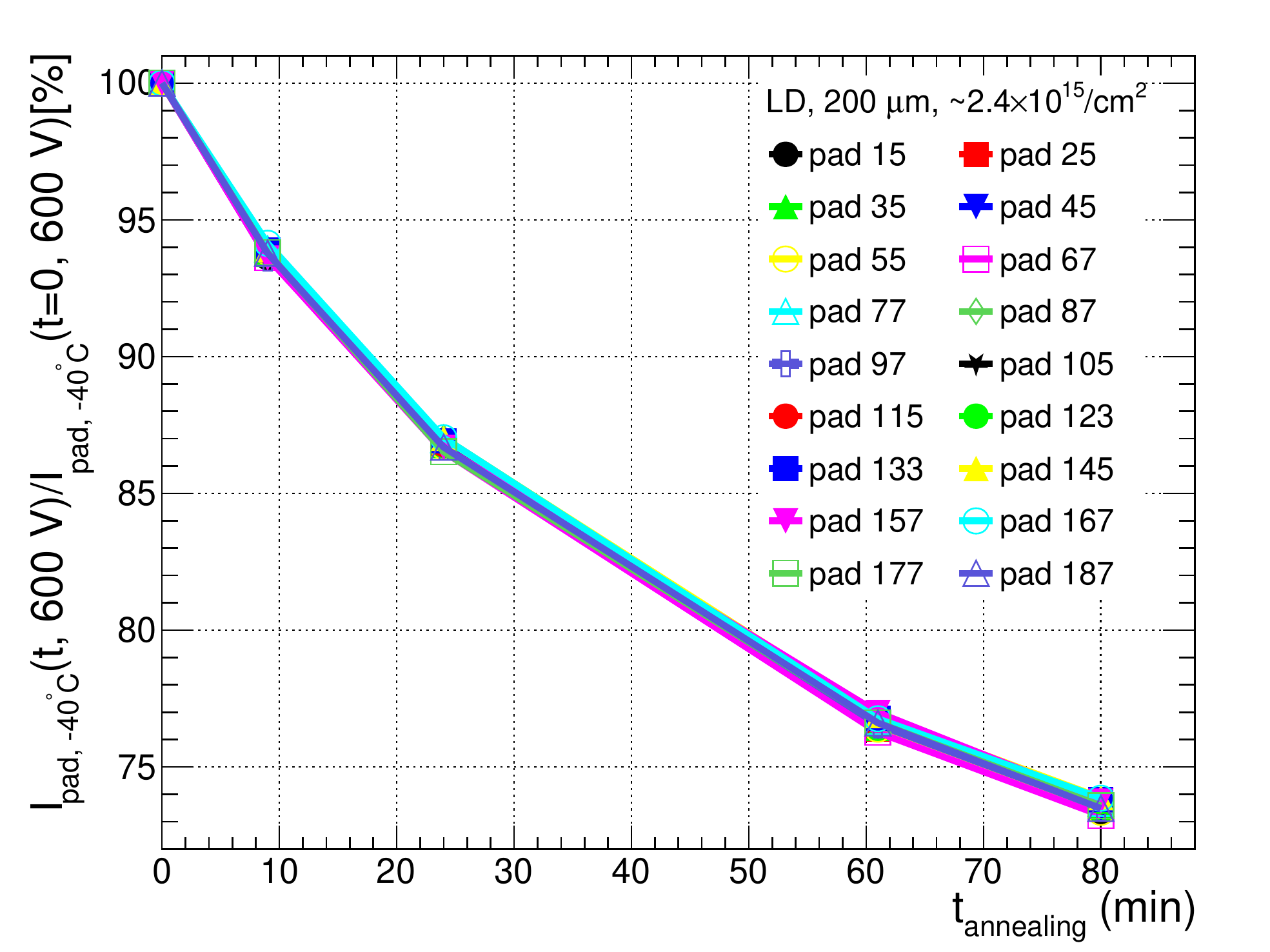}
		\subcaption{
		}
		\label{plot:annealing_current}
	\end{subfigure}    

	\caption{
		(a) IV-curves of a representative full hexagonal pad for different annealing durations for a \SI{200}{\micro\metre} low-density prototype sensor irradiated to approximately 2.4$\times 10^{15}~\neqcm$.
        (b) Decrease of the per-pad leakage current ($U_\text{bias}=\SI{600}{\volt}$) as a function of the annealing time ($\text{t}_\text{annealing}$) at \SI{60}{\celsius}.
	}
\end{figure}
\begin{figure}
	\captionsetup[subfigure]{aboveskip=-1pt,belowskip=-1pt}
	\centering
    \includegraphics[width=0.69\textwidth]{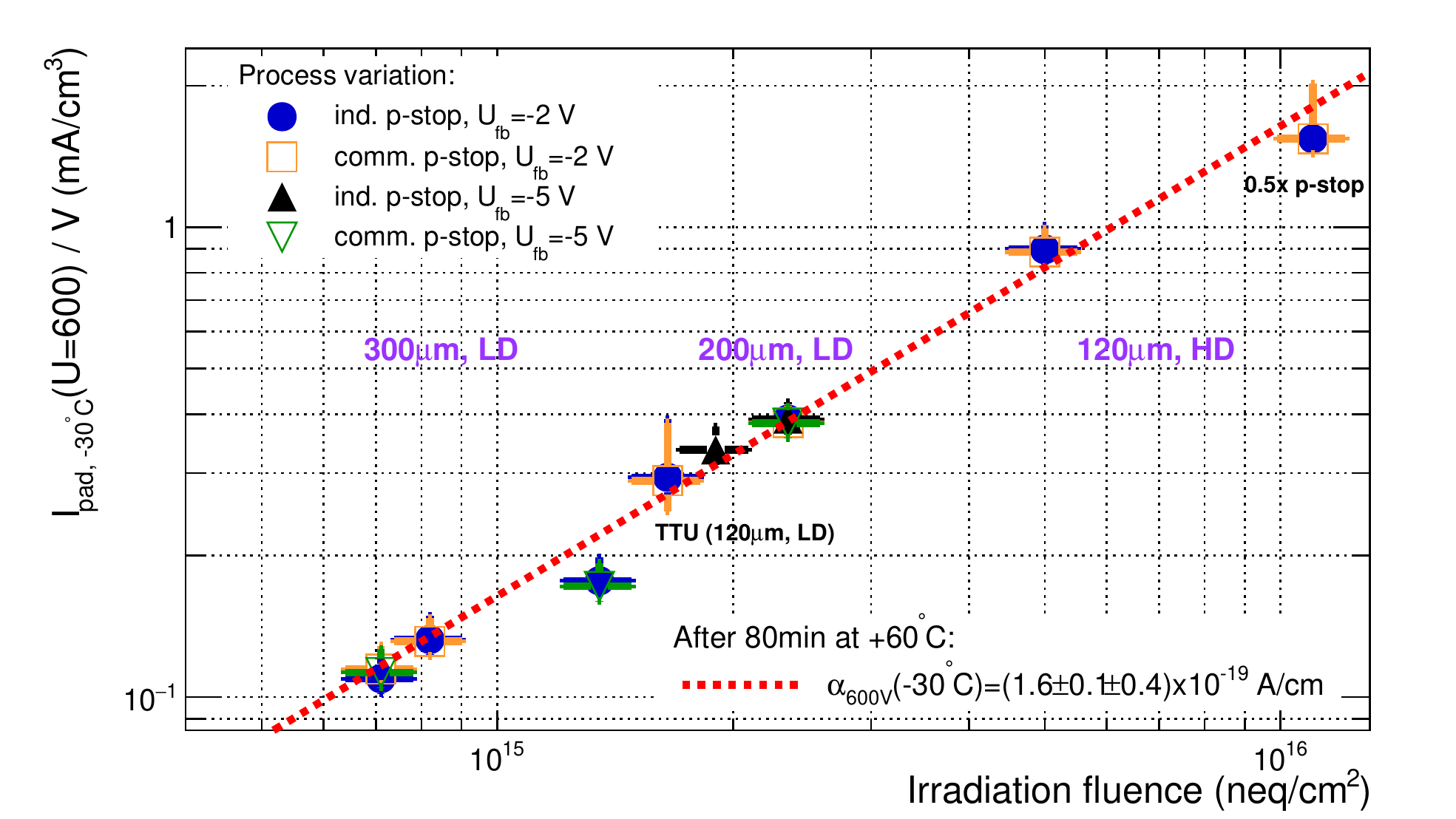}
	\caption{
		Volume-normalised per-pad leakage currents interpolated to an effective bias voltage of \SI{600}{\volt} as a function of the fluence that they were exposed to.
        The quoted currents were measured after additional annealing at \SI{60}{\celsius}, and are scaled to the CE's planned operation temperature of \SI{-30}{\celsius}.
		"0.5x p-stop" denotes HD sensors with half the p-stop concentration.
		"TTU" denotes the measurements of \SI{120}{\micro\meter} HD sensors conducted with a analogous setup to the one described in~\ref{subsec:setup_alps}, at Texas Tech University. 
		Those exhibit overall consistency with the measurements conducted at CERN.
		}
	\label{plot:alpha_600}
\end{figure}
\begin{figure}
	\captionsetup[subfigure]{aboveskip=-1pt,belowskip=-1pt}
	\centering
	\begin{subfigure}[b]{0.32\textwidth}
		\includegraphics[width=0.999\textwidth]{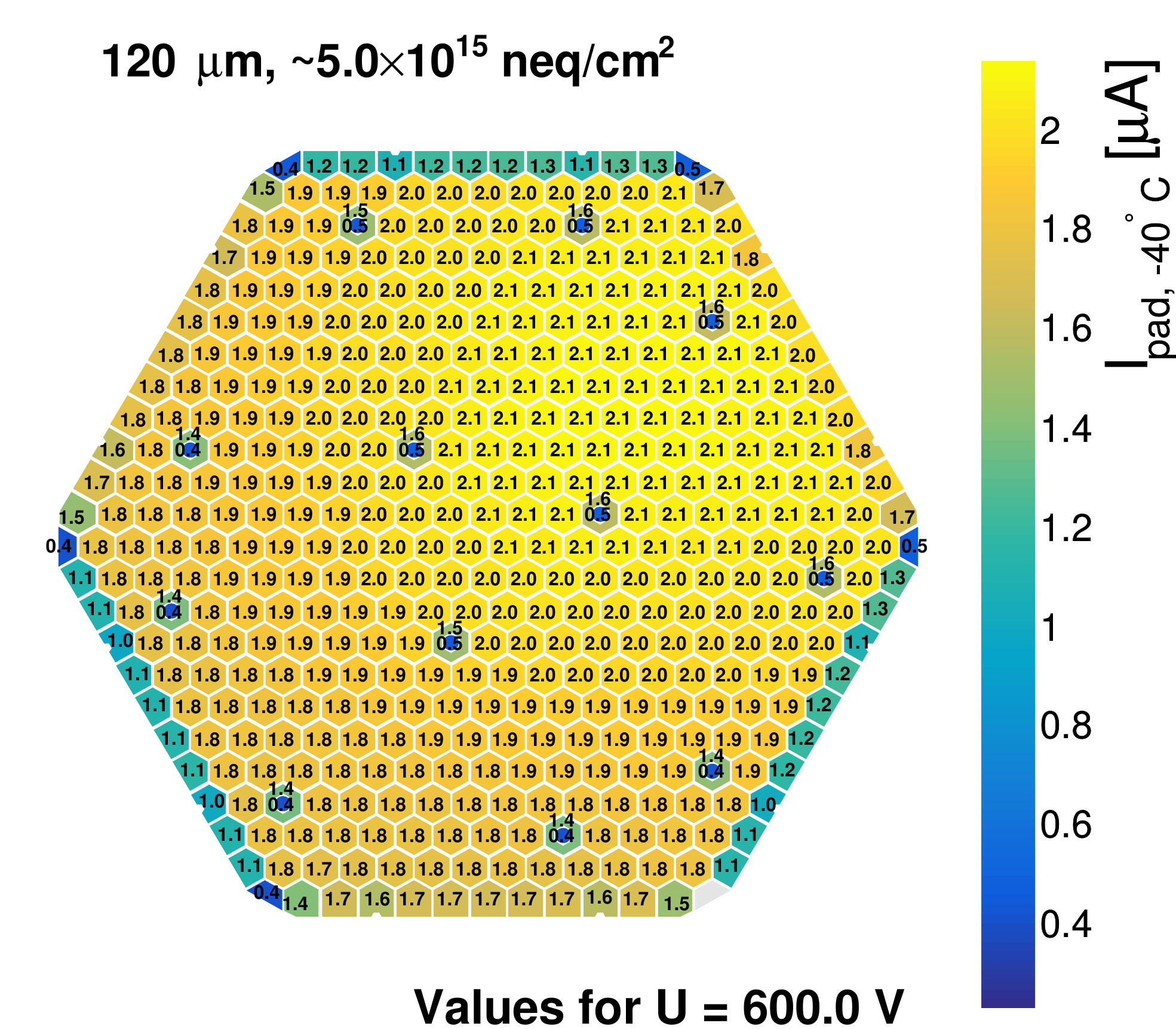}
		\subcaption{
		}
		\label{plot:iv_hexplot_3009}
	\end{subfigure}
	\hfill
	\begin{subfigure}[b]{0.32\textwidth}
		\includegraphics[width=0.999\textwidth]{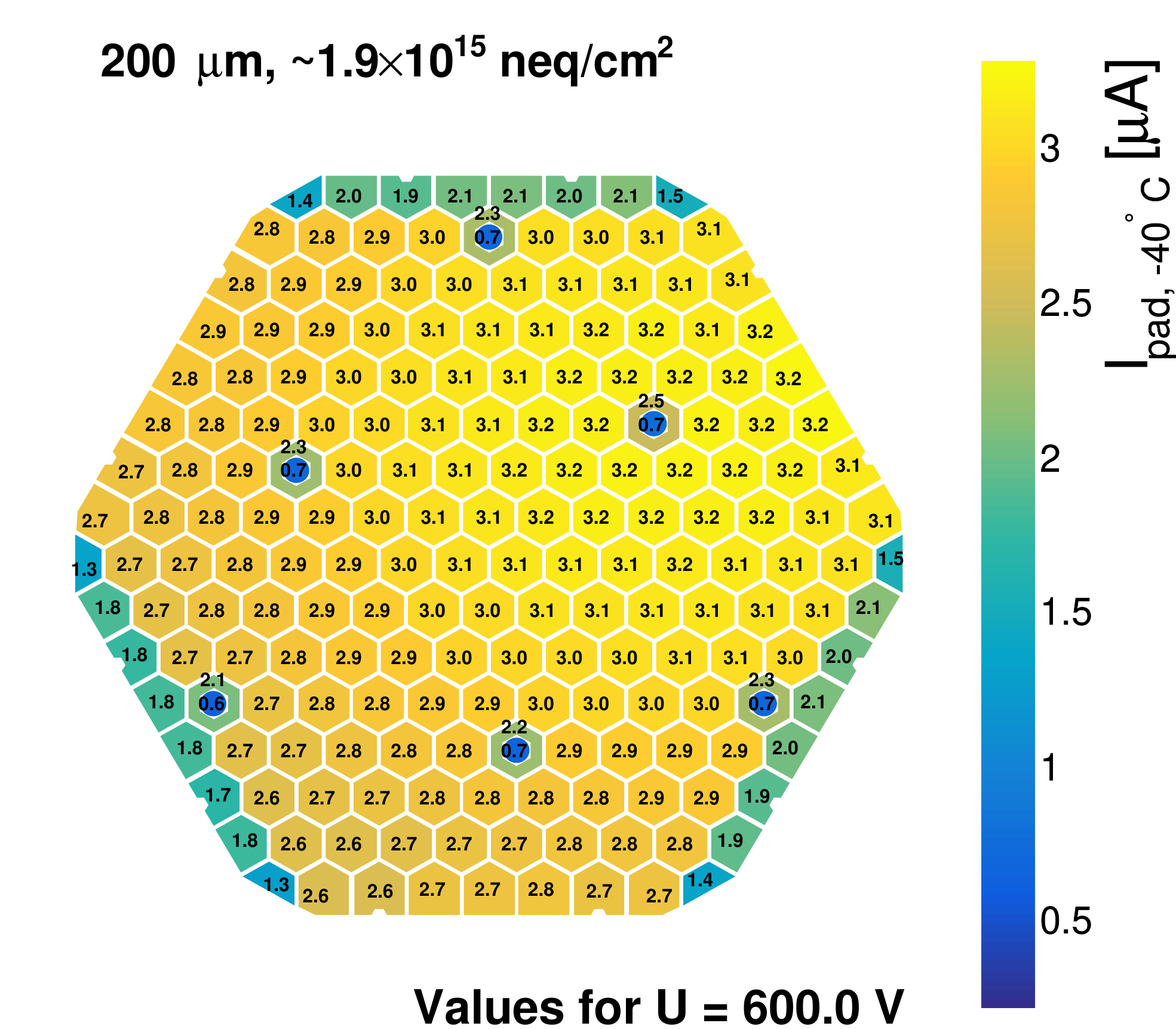}
		\subcaption{
		}
		\label{plot:iv_hexplot_0541_04}
	\end{subfigure}
	\hfill	
	\begin{subfigure}[b]{0.32\textwidth}
		\includegraphics[width=0.999\textwidth]{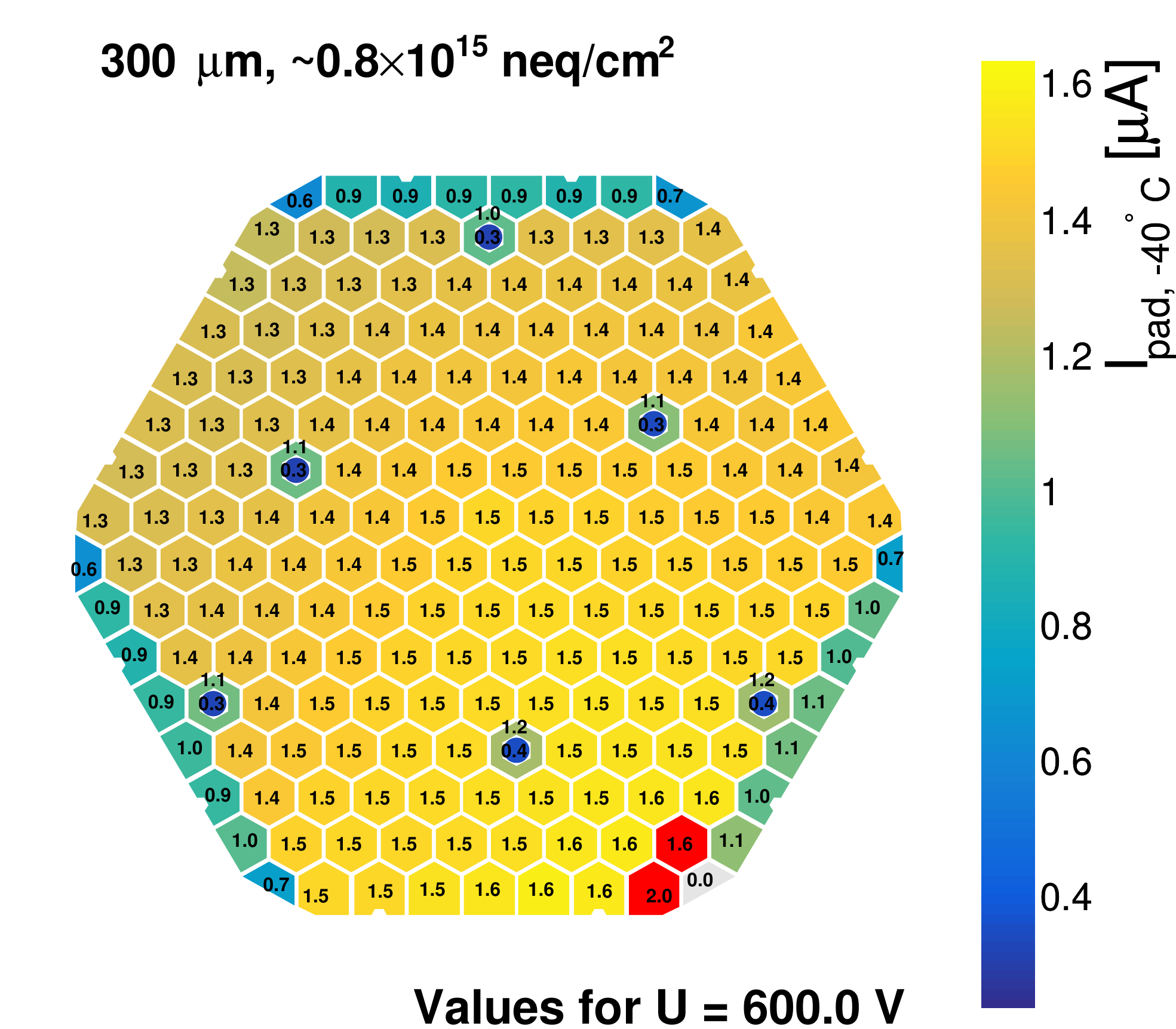}
		\subcaption{
		}
		\label{plot:iv_hexplot_1013}
	\end{subfigure}
    \hfill
	\begin{subfigure}[b]{0.32\textwidth}
		\includegraphics[width=0.999\textwidth]{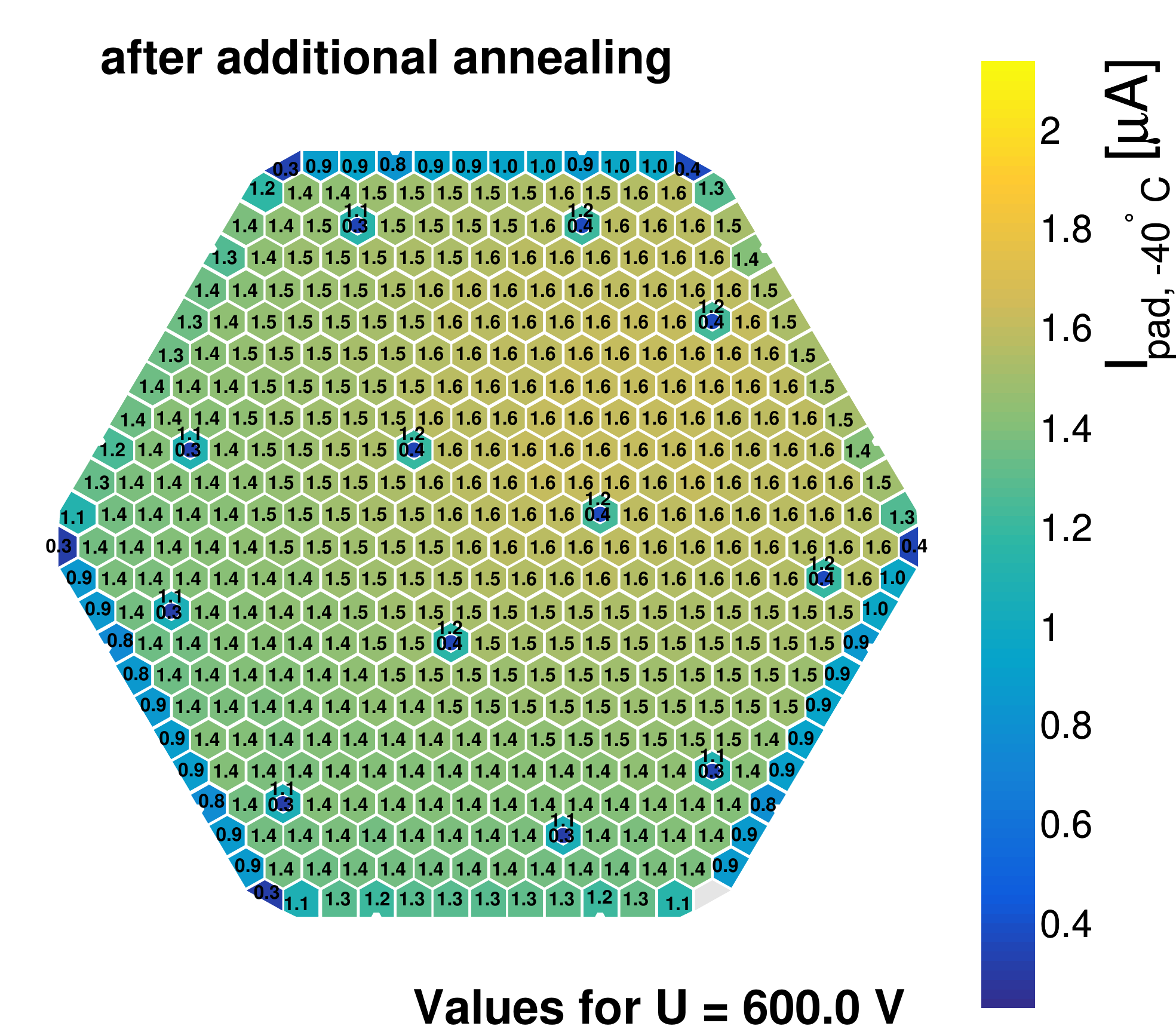}
		\subcaption{
		}
		\label{plot:iv_hexplot_3009_annealed}
	\end{subfigure}
	\hfill
	\begin{subfigure}[b]{0.32\textwidth}
		\includegraphics[width=0.999\textwidth]{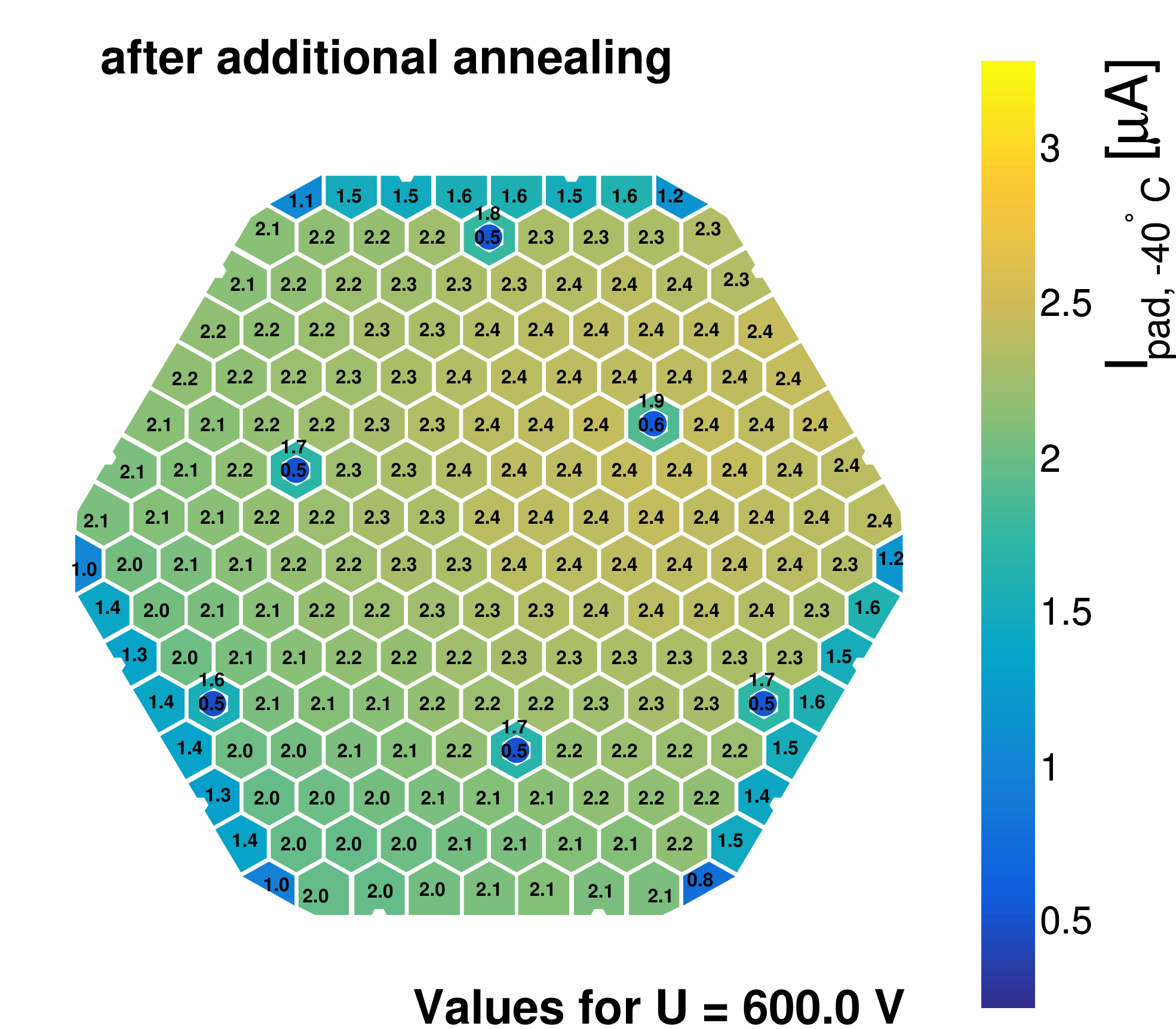}
		\subcaption{
		}
		\label{plot:iv_hexplot_0541_04_annealed}
	\end{subfigure}
	\hfill	
	\begin{subfigure}[b]{0.32\textwidth}
		\includegraphics[width=0.999\textwidth]{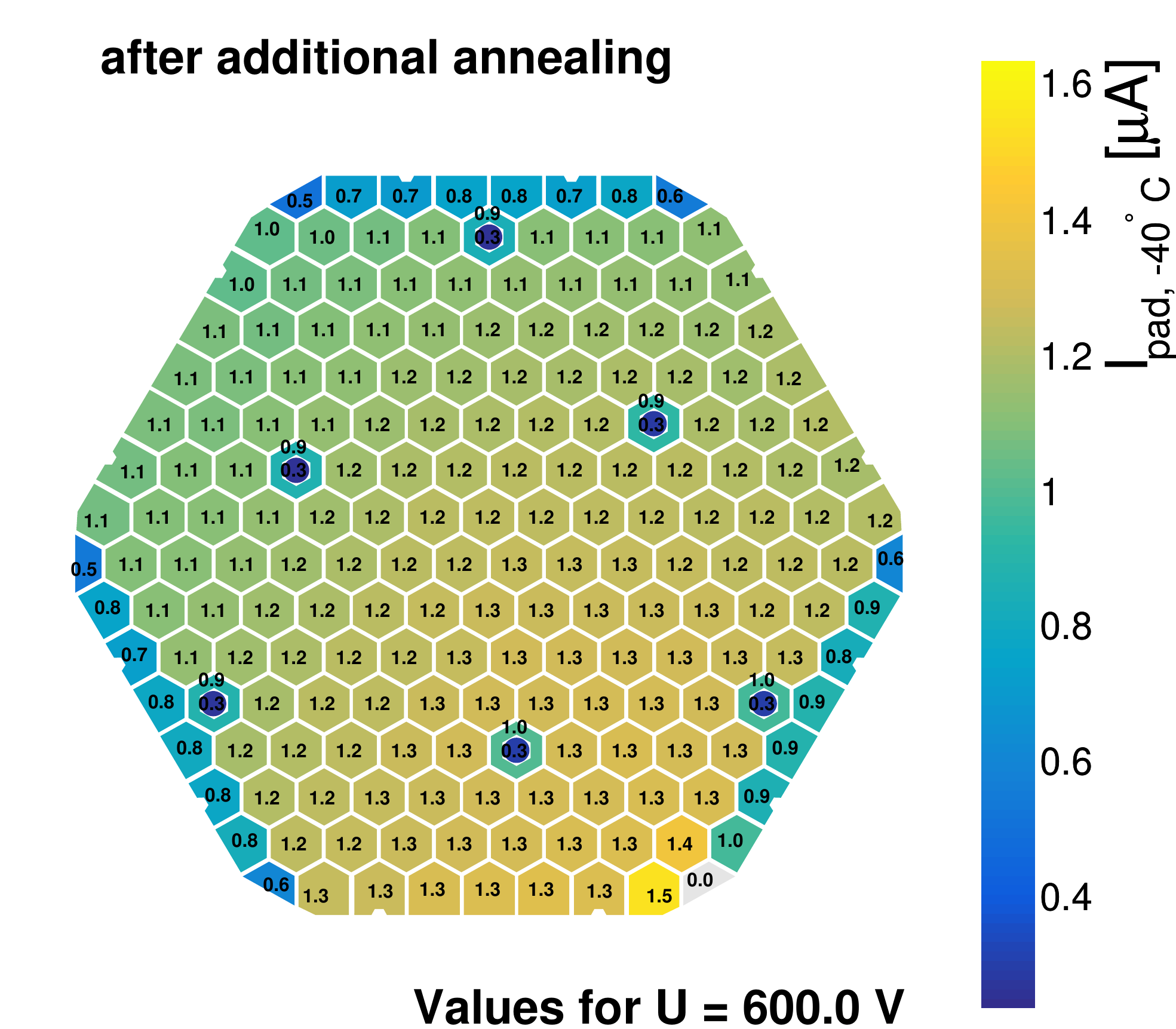}
		\subcaption{
		}
		\label{plot:iv_hexplot_1013_annealed}
	\end{subfigure}    
	\caption{
		Per-pad leakage currents interpolated to an effective bias voltage of \SI{600}{\volt} for three representative sensors from different irradiation rounds before (a-c) and after additional annealing (d-f).
		Red- or white-colored edge pads correspond to well-understood measurement effects, e.g. insufficient contact between the pogo pins and the pads.
		}
	\label{plot:iv_hexplot}
\end{figure}

The total leakage current, interpreted as the dark current of a full silicon sensor, is defined as the current flowing through the high-voltage resistor R$_\text{HV}$ (cf. Figure 2 in~\cite{pitters:array2019}).
As it is shown for two prototype sensors in \ref{plot:tot_IV_good}, the total leakage current did not break down and stayed well below the ARRAY system's compliance of \SI{2}{\milli\ampere} for the irradiated sensors discussed in this paper.
Irreversible discharges are undesired but occurred for a handful of sensors where the total leakage current suddenly increased and exceeded the \SI{2}{\milli\ampere} limitation.
One half of those instances could be traced back to mechanical damages, e.g.$~$induced during the transport or sensor handling, and the other half hinted at the presence of a minor flaw in the CE silicon sensor design.
The latter ultimately lead to a design modification with which the risk of discharges in the future should be minimised\footnote{More than 50 prototype sensors with the improved design have been characterized with voltages up to \SI{850}{\volt} in the meantime. None have shown discharges thus far.}.
Results of the affected sensors are not discussed further in this paper.\newline
\ref{plot:pad_IV_channels} shows the per-pad leakage current as a function of the effective bias voltage for adjacent pads of various sizes on a low density sensor irradiated up to $\sim 1.9\times 10^{15}~\neqcm$.
The data demonstrate that, in good approximation, the leakage current of a pad scales with its volume.
The relative increase from \SI{600}{\volt} to \SI{800}{\volt} remains well below \SI{150}{\percent} meeting the vendor specifications.\newline
The IV curves after different annealing times are shown for a representative pad in \ref{plot:annealing_IV}.
For bias voltages beyond full depletion, per-pad leakage currents at \SI{600}{\volt} are reduced systematically by \SI{25}{\percent} after \SI{80}{\minute} at \SI{60}{\celsius}, cf.~\ref{plot:annealing_current}.
Given the simultaneous reduction in the depletion voltage, cf.~\ref{plot:annealing_Vdep}, leakage currents around full depletion are even further reduced.\newline
The expected relationship between the per-pad leakage current density and the fluence was observed.
The proportionality for current densities at a bias voltage of \SI{600}{\volt}, which is well above full depletion for all investigated sensors (except for the \SI{300}{\micro\metre} sensors irradiated to $13.5\times 10^{14}\neqcm$), and extrapolated to \SI{-30}{\celsius} is displayed in \ref{plot:alpha_600}.
As expected~\cite{MOLL199987}, the current-related damage rate ($\alpha$) is found to be independent on the tested material properties investigated in this work.
The numerical result for $\alpha$ scaled to room temperature is $\alpha_\text{600V}(\SI{+20}{\celsius})=\left(3.2\pm 0.2\pm 0.6\right)\times 10^{-17}~$A/cm.
The central value is about \SI{20}{\percent} less than what is reported in Ref.~\cite{moll:SiDamages}.
A possible explanation could be a systematic overestimation of the fluence at the RINSC irradiation facility.
Follow-up studies in this regard are anticipated in the mid-term future.\newline 
After correction for the chuck temperature non-uniformity, per-pad leakage current densities across a sensor at a fixed voltage still vary by $\sim 10~\%)$.
The associated current profiles are present both before (cf. \ref{plot:iv_hexplot_3009,plot:iv_hexplot_0541_04,plot:iv_hexplot_1013}) and after additional annealing (cf. \ref{plot:iv_hexplot_3009_annealed,plot:iv_hexplot_0541_04_annealed,plot:iv_hexplot_1013_annealed}).
Furthermore, they are consistent between sensors that had been irradiated simultaneously in the same puck.
Among others, a plausible explanation for this observation could be the presence of a fluence profile within the beam port at RINSC that can be approximated as a Gaussian with a width of $\sigma\sim\SI{10}{\centi\metre}$.

\subsection{Capacitance and Depletion Voltage}
\label{subsec:Udep}
\begin{figure}
	\captionsetup[subfigure]{aboveskip=-1pt,belowskip=-1pt}
	\centering

	\begin{subfigure}[b]{0.49\textwidth}
		\includegraphics[width=0.999\textwidth]{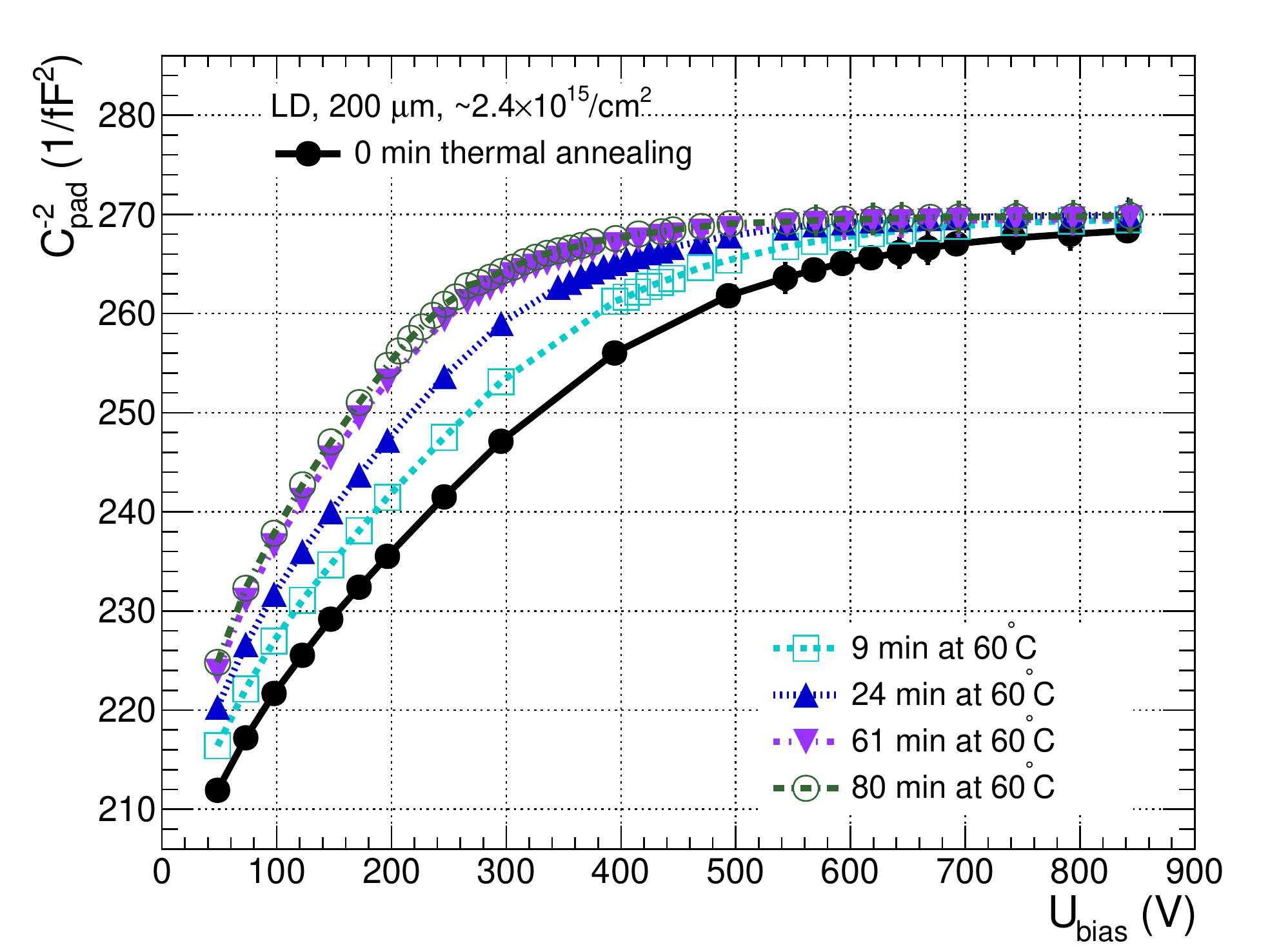}
		\subcaption{
		}
        \label{plot:annealing_CV}
	\end{subfigure}
    \hfill
    \begin{subfigure}[b]{0.49\textwidth}
		\includegraphics[width=0.999\textwidth]{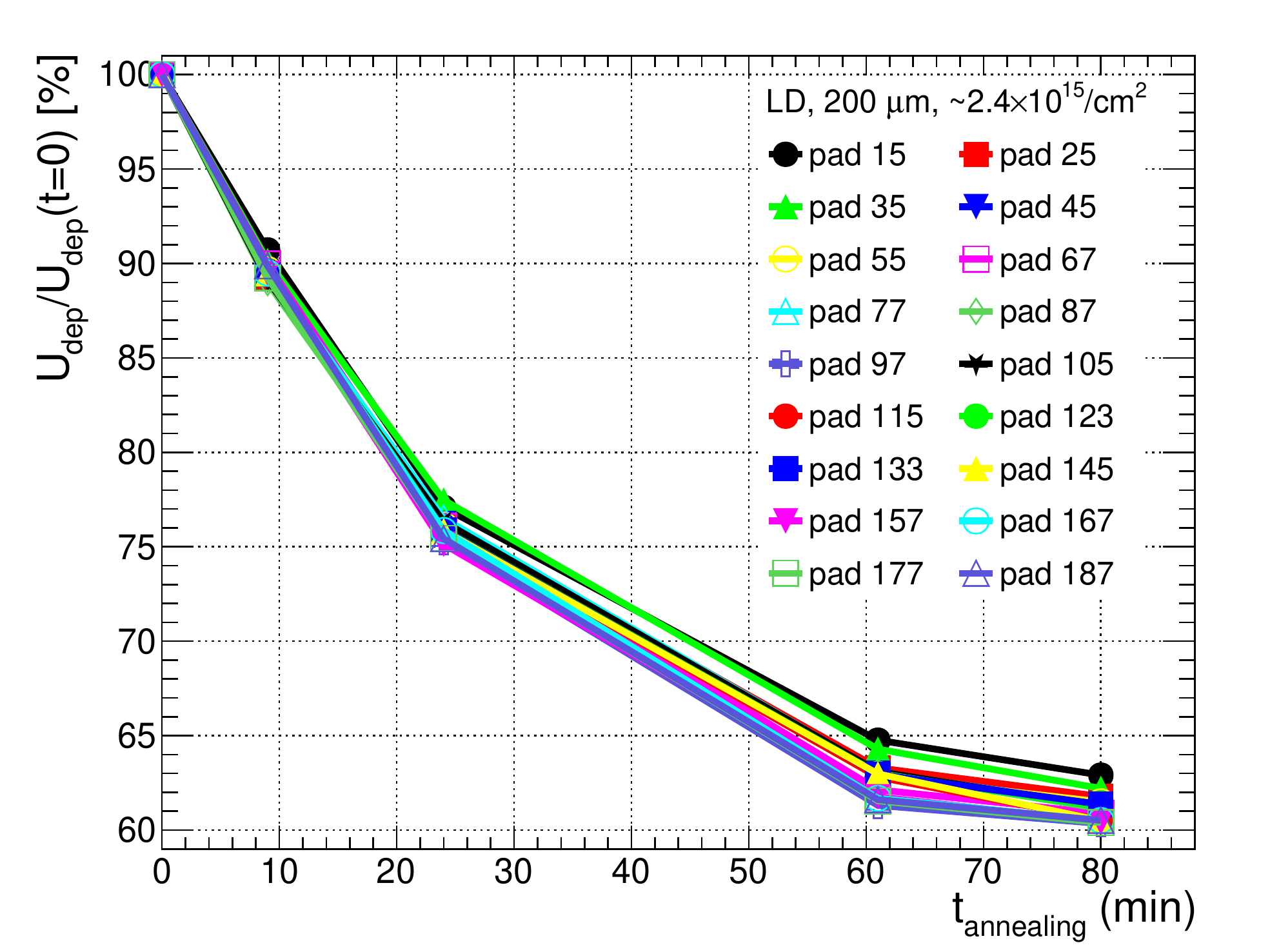}
		\subcaption{
		}		
        \label{plot:annealing_Vdep}
	\end{subfigure}
	\caption{
        (a) Reciprocal-squared capacitance ($C^{-2}_\text{pad}$) as a function the bias voltage of a representative full hexagonal pad for different annealing durations for a \SI{200}{\micro\metre} low-density prototype sensor irradiated to approximately 2.4$\times 10^{15}~\neqcm$.   
		(b) Relative decrease of the depletion voltage estimate ($U_\text{dep}$) as a function of the additional annealing time ($\text{t}_\text{annealing}$) at \SI{60}{\celsius} for a subset of full pads.
	}
\end{figure}

\begin{figure}
	\captionsetup[subfigure]{aboveskip=-1pt,belowskip=-1pt}
	\centering
	\begin{subfigure}[b]{0.49\textwidth}
		\includegraphics[width=0.999\textwidth]{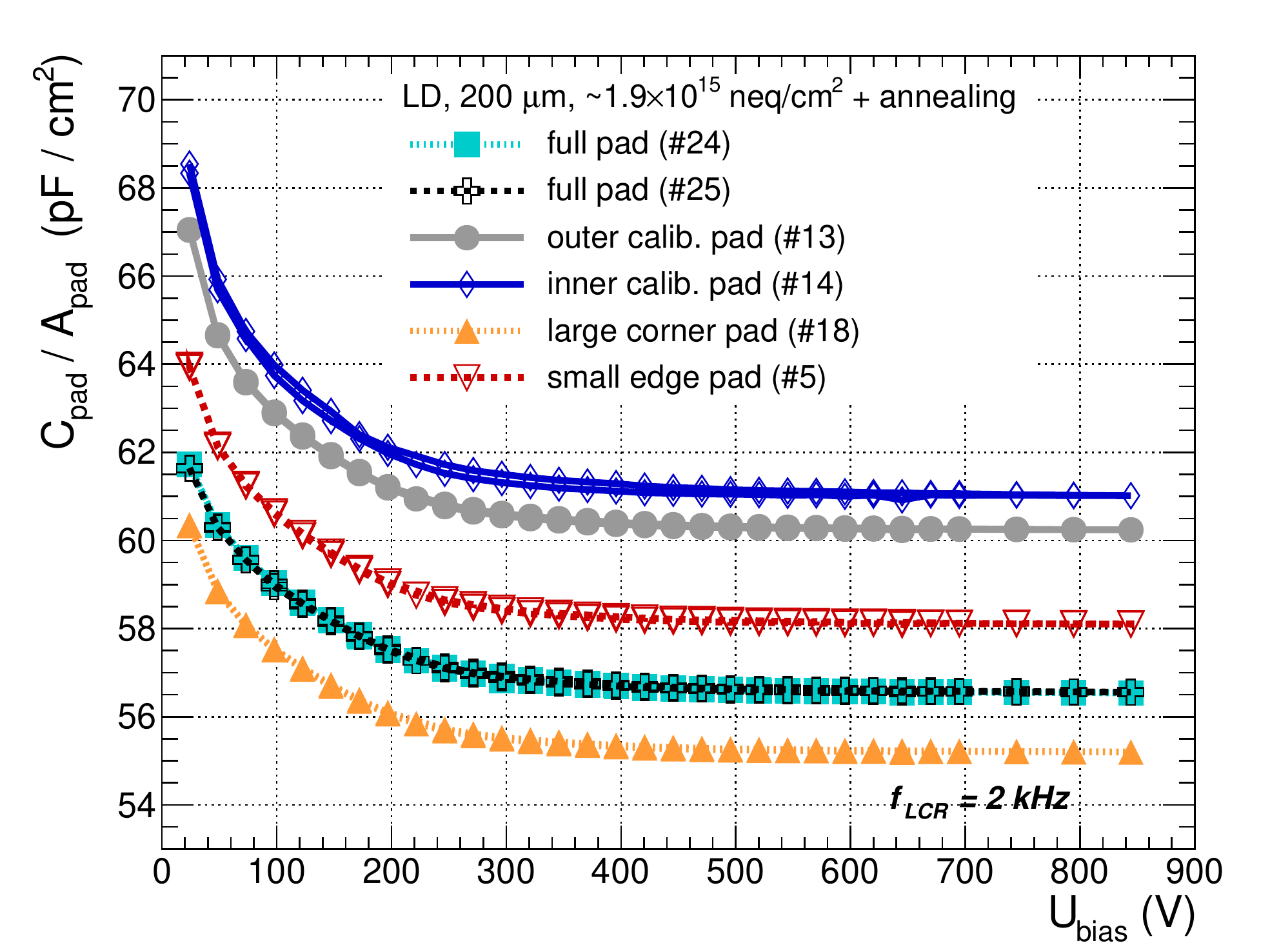}
		\subcaption{
		}
		\label{plot:pad_CV_channels}
	\end{subfigure}
	\hfill
	\begin{subfigure}[b]{0.49\textwidth}
		\includegraphics[width=0.999\textwidth]{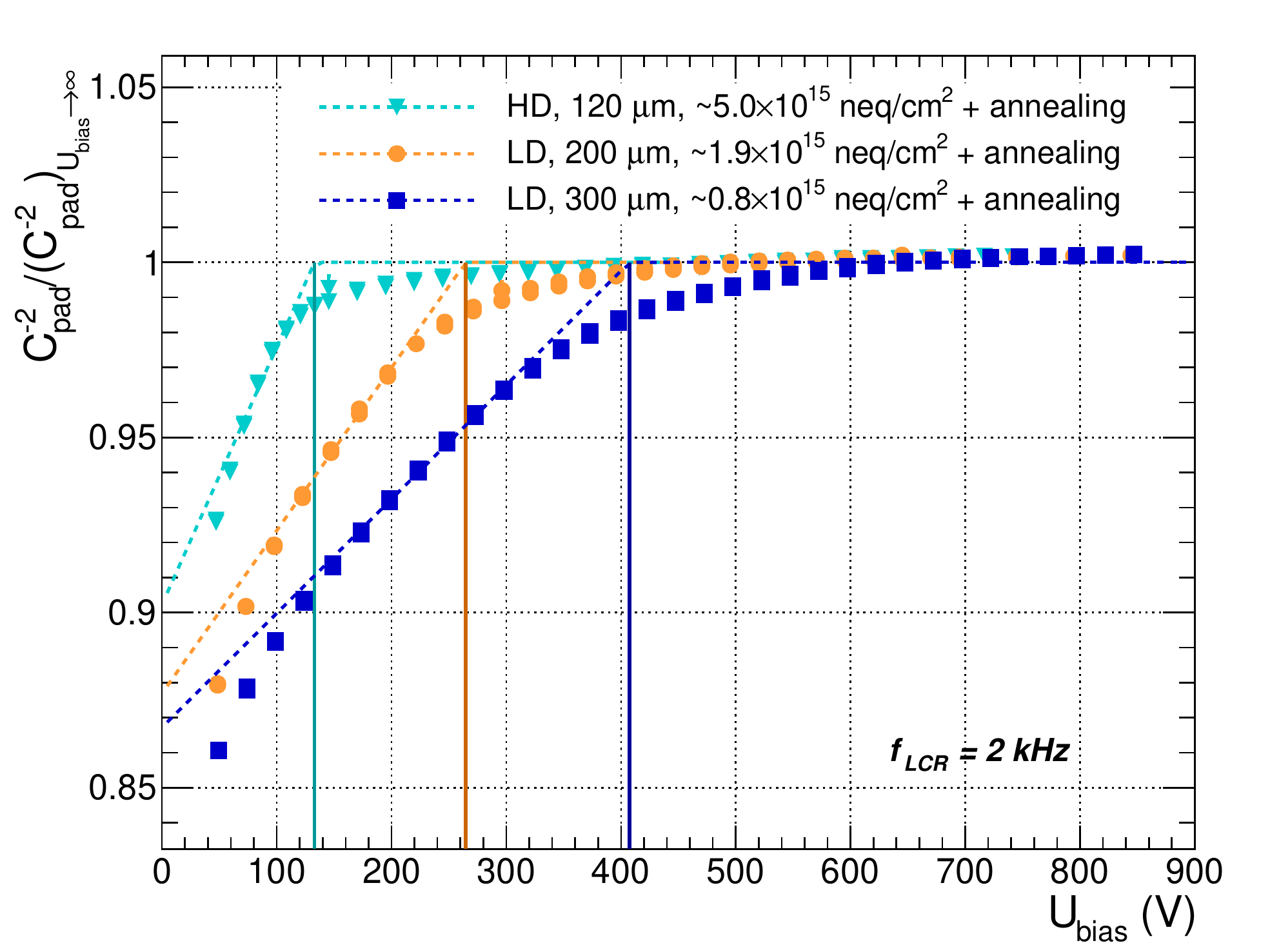}
		\subcaption{
		}
		\label{plot:pad_invCV_sensor}
	\end{subfigure}
	\caption{
		(a) Area-normalised capacitances as a function of the bias voltage for different pads with different geometries on a example LD sensor.
		(b) Normalised squared-inverse capacitances as a function of the bias voltage to estimate the sensor depletion voltage of a central pad from different irradiation rounds.
		Full depletion was achieved at bias voltages around \SI{40}{\volt} for \SI{120}{\micro\metre} sensors, \SI{120}{\volt} for \SI{200}{\micro\metre} sensors, and \SI{280}{\volt} for the \SI{300}{\micro\meter} sensors before irradiation. 
	}
\end{figure}

The measured per-channel impedance is open-corrected by subtracting the impedance of the ARRAY system.
Subsequently, the sensor pad capacitance is computed assuming an underlying serial connection to the measurement circuitry.
Due to the finite mobility of the sensor defects, the frequency at which the impedance is measured may, in principle, have sizeable impact on the derived capacitance and on the depletion voltages, cf. Ref.~\cite{Li1991}.
The impedance measurement was conducted at $f_\text{LCR}=\SI{2}{\kilo\hertz}$.
Experimental follow-up indicated an impact of \SI{2}{\percent} on the asymptotic capacitance and a \SI{10}{\percent} reduction of the depletion voltage when decreasing $f_\text{LCR}$ to \SI{500}{\hertz}, which have no implication on the following discussion.\newline
The depletion voltage for each pad is assessed from the saturation of its squared reciprocal capacitance ($C^{-2}$) of a pad with respect to the bias voltage ($V$). 
It is defined as the intersection of a straight-line fitted to the rising part of the $C^{-2}$ \emph{vs.} $V$ curve with a line fitted to its plateau.
Prior to additional annealing, such a plateau could not always be reached within the tested bias voltage range, see for example the black line in \ref{plot:annealing_CV}. 
After additional annealing to \SI{80}{\min} at \SI{60}{\celsius}, the estimated depletion voltage has been reduced by about a third with respect to the case without annealing, cf.~\ref{plot:annealing_Vdep}.\newline
The per-pad capacitance after irradiation is found to scale well with the area of the pad, cf.~\ref{plot:pad_CV_channels}.
Deviations from this scaling amount to less than \SI{5}{\percent}, suggesting a slight dependence of the inter-pad capacitances on the pad geometries.\newline
\ref{plot:pad_invCV_sensor} illustrates that the estimated depletion voltages depend primarily on the associated thickness of the depleted zone. 
\begin{figure}
	\captionsetup[subfigure]{aboveskip=-1pt,belowskip=-1pt}
	\centering
	\begin{subfigure}[b]{0.49\textwidth}
		\centering
		\includegraphics[width=0.99\textwidth]{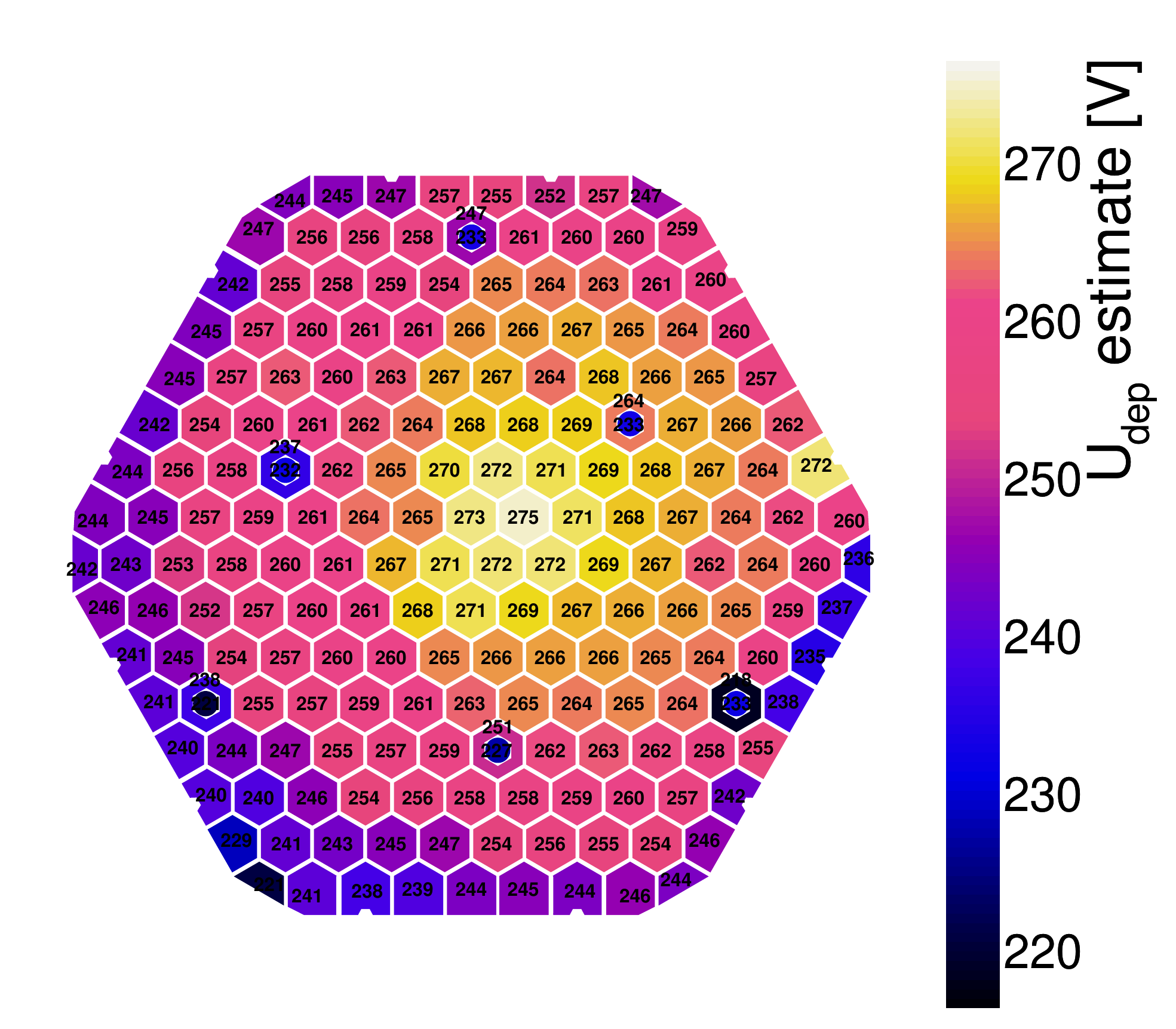}
		\subcaption{
			}
			\label{plot:Vdep_hexplot_0541_04}
	\end{subfigure}
	\hfill
	\begin{subfigure}[b]{0.49\textwidth}
		\centering
		\includegraphics[width=0.999\textwidth]{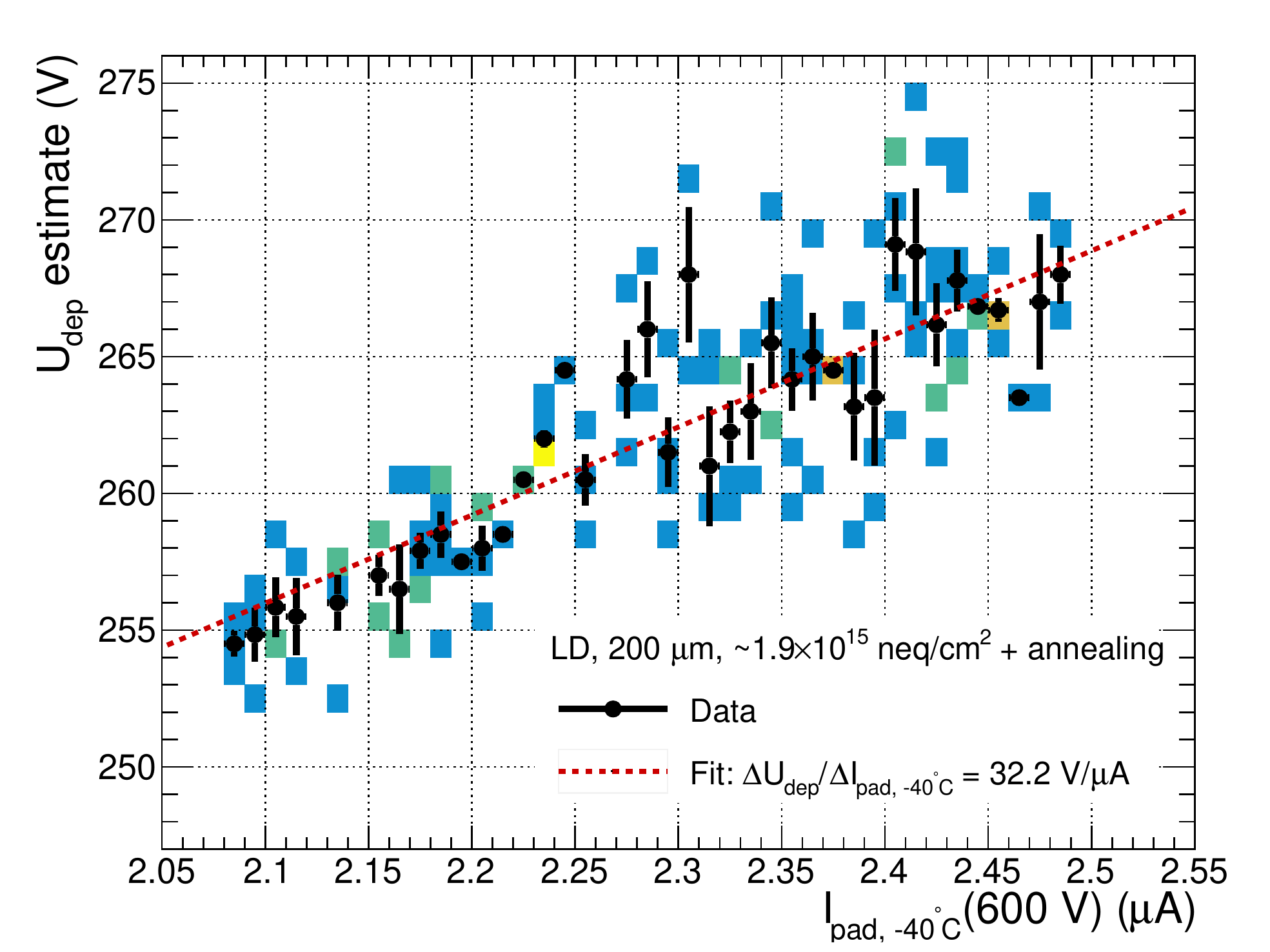}
		\subcaption{
			}
			\label{plot:Vdep_vs_current_5414}
	\end{subfigure}
	\caption{
		(a) Per-pad depletion voltage estimates for a \SI{200}{\micron} LD example sensor irradiated to 1.9$\times 10^{15}~\neqcm$, and 
		(b) their correlation to the per-pad leakage current, used as proxy for the delivered fluence.
	}
\end{figure}
The estimated per-pad depletion voltage across the sensor is shown in~\ref{plot:Vdep_hexplot_0541_04}.
Those estimates exhibit positive correlation with the leakage currents, taken as proxy for the fluence, cf. \ref{plot:Vdep_vs_current_5414}.
In fact, the obtained correlation coefficient from this analysis is positive for all tested sensors where full depletion could be reached.
This can be interpreted as further evidence for the presence of a fluence profile inside the beam port during irradiation.

\subsection{Discussion}
\label{subsec:discussion}
The neutron irradiation induced leakage currents are consistent both qualitatively and quantitatively with previous R$\&$D on irradiated silicon sensors.
They are also found to be independent of the silicon fabrication process (flatband voltage, p-stop layout).
Furthermore, unchanged end-capacitances and reduced depletion voltages after beneficial annealing are qualitatively consistent with previous findings.\newline
The findings presented in this paper with the first prototypes of CE silicon sensors, i.e. hexagonal sensors and fabricated with a new 8'' process, irradiated to fluences corresponding to the end of lifetime of HL-LHC are in full support of the radiation-hardness of their general design.
By implication, RINSC could be qualified as a valid facility for neutron irradiation of 8'' silicon sensors for the first time.
However, more accurate studies on the electrical sensor properties after neutron-irradiation would need to incorporate the systematic uncertainty on the effective annealing during irradiation, that is due to the temperature evolution and its potential lateral profile inside the reactor's beam port.
Similarly, an accurate assessment of the actual fluence at each pad would have to address the neutron flux profile, for whose existence this work provides first evidence.

%% file: content/7_conclusion.tex
\section{Conclusion}
\label{sec:conclusion}
An essential component of the silicon sensors' prototyping for the CMS highly granular endcap calorimeter upgrade (CE) is the experimental verification of the radiation hardness in terms of their electrical properties, i.e. leakage currents, capacitances and depletion voltages.
RINSC is one of the few locations world-wide that offers the infrastructure for the irradiation of large structures compatible with CE's 8''-wide silicon pad sensors.
Given that parts of the necessary infrastructure have not been used in this experimental capacity, the irradiation was non-trivial and specific preparations had to be made.
After irradiation, the probe- and switch card-based ARRAY system~\cite{pitters:array2019} was used for the first time for the electrical characterisation of neutron-irradiated silicon sensors at CERN and at Texas Tech University.
In order to protect the test system from currents that would exceed its specifications, the neutron-irradiated sensors were cooled down to \SI{-40}{\celsius} during the tests.\newline
The preparation and execution of the neutron-irradiation at RINSC and the electrical characterisation of the irradiated sensors at cold temperatures using the ARRAY system was explained in this paper.
Evidence for specific properties of the RINSC irradiation facility such as a non-uniform flux profile across the 8'' wafer and for non-negligible sensor annealing due to insufficient cooling during irradiation was reported.\newline
For the planned CMS CE upgrade, the findings in this work are encouraging as they confirm the overall expected radiation hardness of the prototype silicon sensors.
In particular, their leakage current densities are found to scale proportionally with the fluence, independent of the properties of the tested production process variations.
It is reconfirmed that annealing the sensors up to \SI{80}{\minute} at \SI{60}{\celsius} has beneficial impact on the electrical properties by lowering the dark current and the depletion voltages by $\mathcal{O}\left(10~\%\right)$.
Despite the tested sensors being prototypes, the insights from their characterisation can be considered representative for the final version.
Therefore, the results reported in this work may serve as reference for the expected electrical performance and degradation of CE's silicon sensors towards the end of their lifetime at HL-LHC.

%% file: content/acknowledgments.tex
\acknowledgments
We thank the staff at the Rhode Island Nuclear Science Center for their support during the preparation and execution of the neutron irradiation of the CE silicon pad sensor prototypes.
The EP-DT and the former EP-LCD group at CERN have developed essential infrastructure, such as the ARRAY system including the associated data acquisition software, and have co-financed the acquisition of the utilised cold-chuck probe station at CERN.
Without their input, this R$\&$D milestone towards the realisation of this novel calorimeter would not have been possible. 
We are thankful for the technical and administrative support at CERN and at other CMS institutes and thank the staffs for their contributions to the success of the CMS effort. 
We acknowledge the enduring support provided by the following funding agencies and laboratories: BMBWF and FWF (Austria); CERN; CAS, MoST, and NSFC (China); MSES and CSF (Croatia); CEA, CNRS/IN2P3 and P2IO LabEx (ANR- 10-LABX-0038) (France); SRNSF (Georgia); BMBF, DFG, and HGF (Germany); GSRT (Greece); DAE and DST (India); MES (Latvia); MOE and UM (Malaysia); MOS (Montenegro); PAEC (Pakistan); FCT (Portugal); JINR (Dubna); MON, RosAtom, RAS, RFBR, and NRC KI (Russia); MoST (Taipei); ThEP Center, IPST, STAR, and NSTDA (Thailand); TUBITAK and TENMAK (Turkey); STFC (United Kingdom); and DOE (USA).

%% file: content/appendix/irradiation_rounds.tex
\section{Irradiation Rounds and Fluences}
\label{appendix:irrad_rounds}
See~\ref{table:irrads}.
\begin{table}[h]
	\centering
	\caption{Irradiation duration and estimated integrated fluences for the neutron-irradiation at RINSC of CE prototype silicon sensors discussed in this work.
	A comparison to leakage currents of silicon test structures irradiated at the JSI TRIGA reactor~\cite{Radulovic:2284353} indicate an uncertainty on the fluence estimates of about 20$~\%$.
	}
	\label{table:irrads}	
	\begin{tabular}{c|ccccc}
		\textbf{Date} & \textbf{No. sensors} & \textbf{Thickness} & \textbf{Duration} & \textbf{Fluence ($10^{14}~\neqcm$)} & \textbf{Reference} \\
		\hline
		26 Aug 20 & 4 & 300$~\mu$m & \SI{13}{\minute} & 7.1 & Si diodes \\
		20 Oct 20 & 4 & 120$~\mu$m & \SI{180}{\minute} & 110.0 & Fe foils \\
		28 Jan 21 & 4 & 200$~\mu$m & \SI{38}{\minute} & 23.5 & Si diodes \\
		11 Feb 21 & 4 & 120$~\mu$m & \SI{38}{\minute} & 16.5 & Si diodes \\
		01 March 21 & 4 & 300$~\mu$m & \SI{23}{\minute} & 13.5 & Si diodes \\
		11 March 21 & 4 & 120$~\mu$m & \SI{76}{\minute} & 50.0 & Fe foils \\
		15 April 21 & 4 & 300$~\mu$m & \SI{15}{\minute} & 8.2 & Si diodes \\
		06 May 21 & 2 & 200$~\mu$m & \SI{38}{\minute} & 19.0 & Si diodes \\
		\hline
	\end{tabular}
\end{table}

%% file: content/appendix/chuck_temperature.tex
\section{Temperature Uniformity of the Systems att C200-40 Cold Chuck}
\label{appendix:chuck_temp}
\begin{figure}
	\centering
	\includegraphics[width=0.69\textwidth]{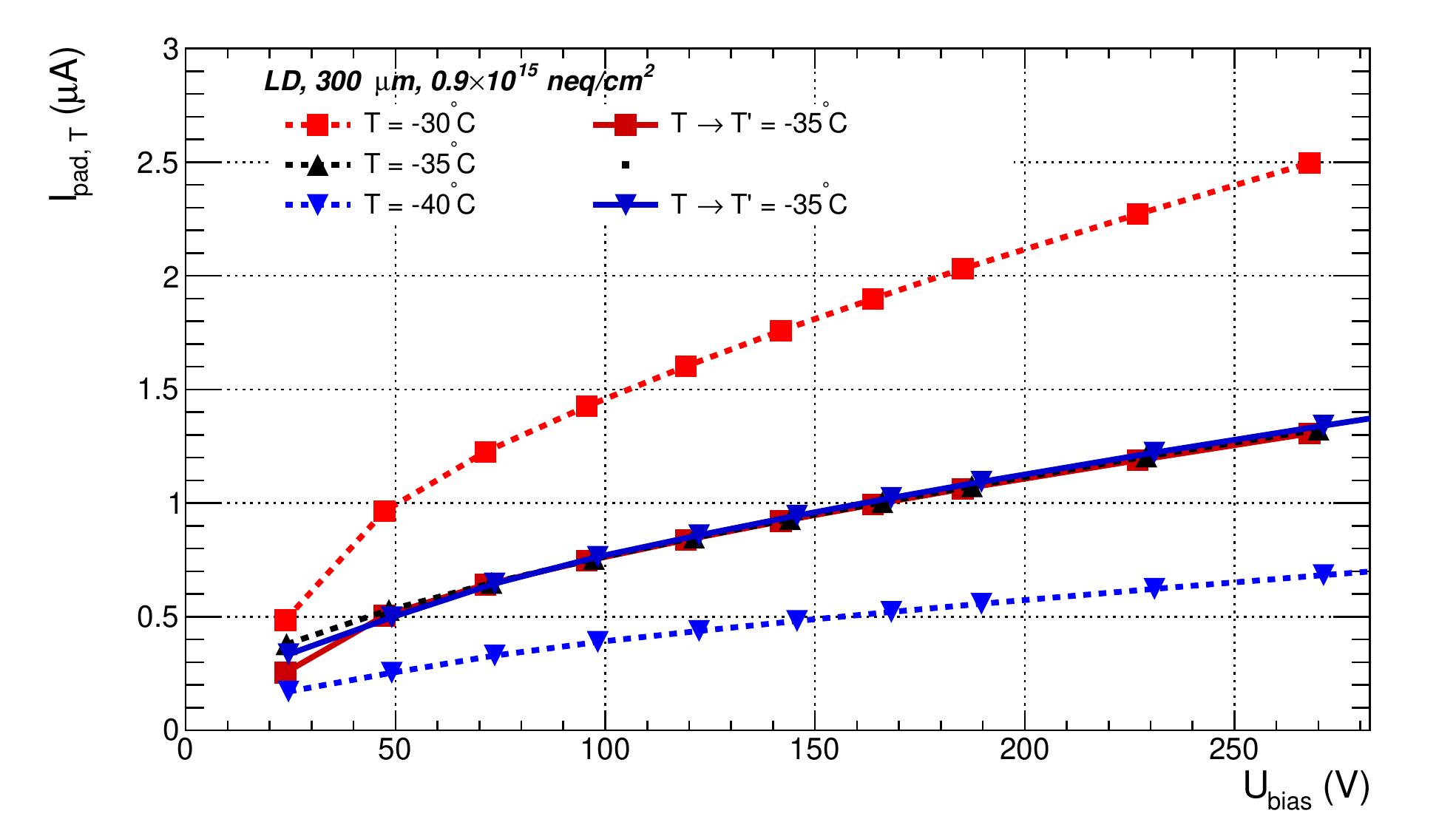}
	\caption{
		Leakage currents as a function of the bias voltage for an example sensor measured at \SI{-40}{\celsius}, \SI{-35}{\celsius}, \SI{-30}{\celsius}, and scaled to \SI{-35}{\celsius} using$~$\ref{eq:temp_scaling}.
		}
	\label{plot:iv_tempscaling}
	\end{figure}

\begin{figure}
	\captionsetup[subfigure]{aboveskip=-1pt,belowskip=-1pt}
	\centering
	\begin{subfigure}[b]{0.32\textwidth}
		\includegraphics[width=0.999\textwidth]{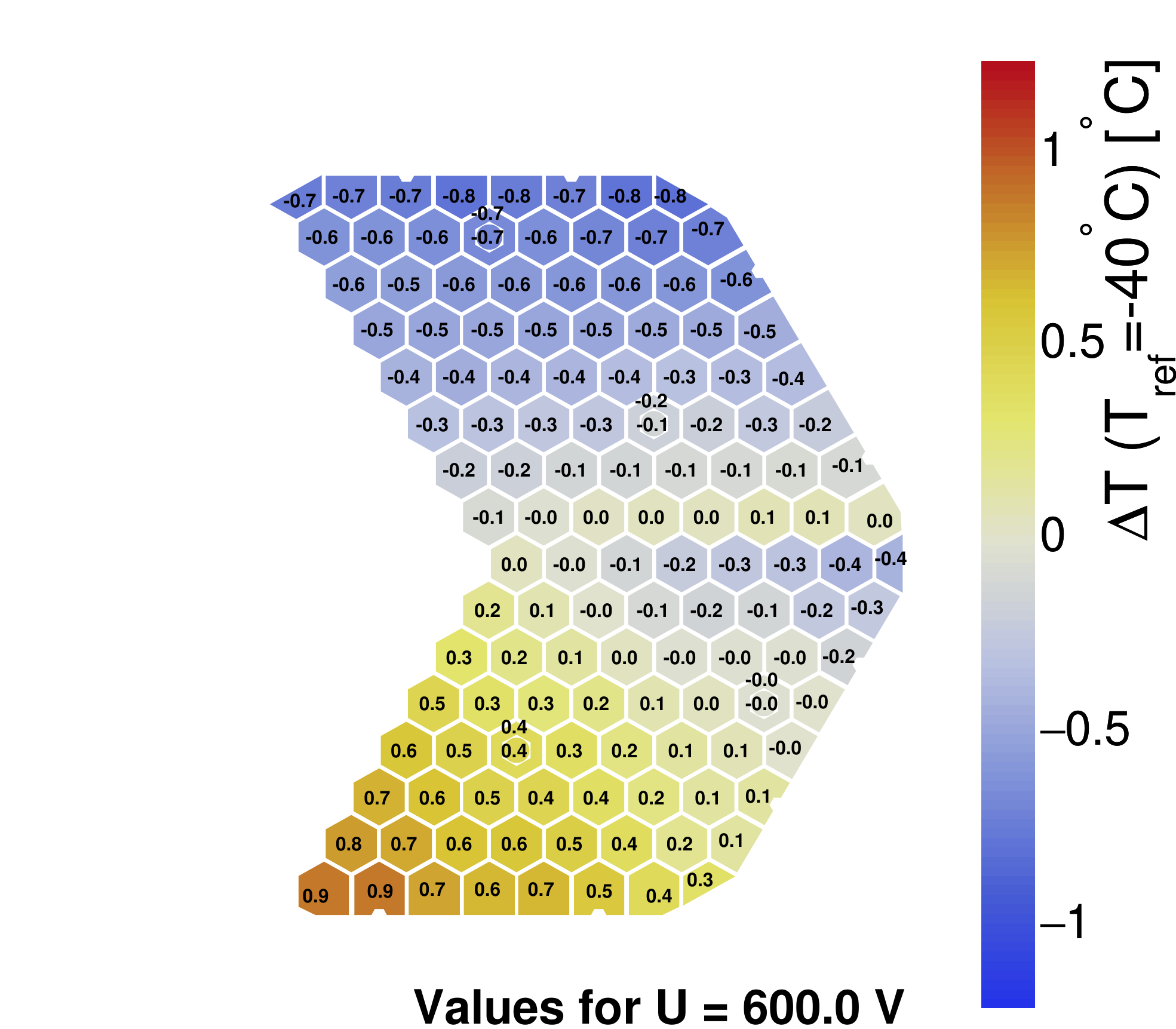}
		\subcaption{
		}
		\label{plot:chucktemp_before}
	\end{subfigure}
	\hfill
	\begin{subfigure}[b]{0.32\textwidth}
		\includegraphics[width=0.999\textwidth]{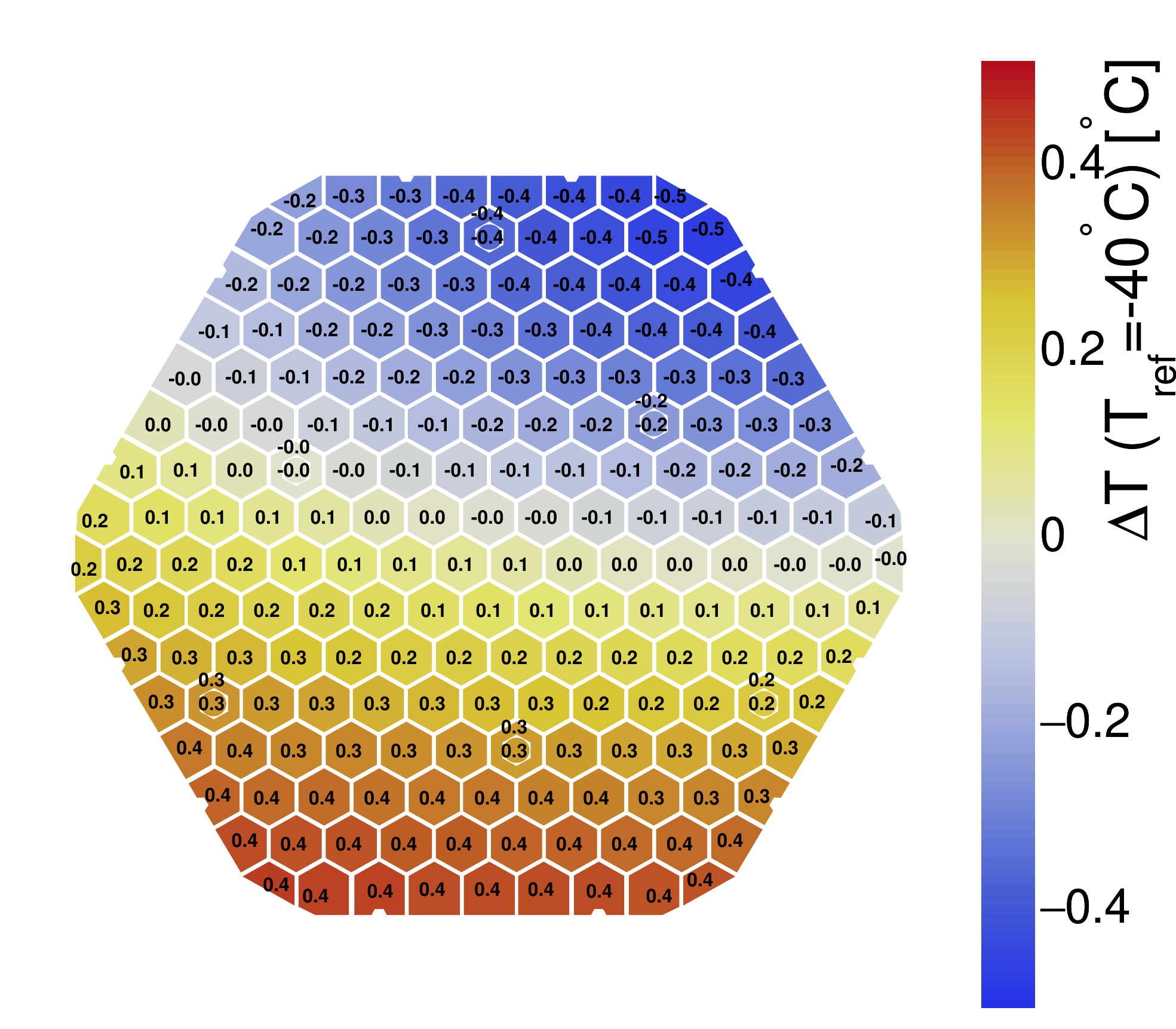}
		\subcaption{
		}
		\label{plot:chucktemp_correction}
	\end{subfigure}
	\hfill
	\begin{subfigure}[b]{0.32\textwidth}
		\includegraphics[width=0.999\textwidth]{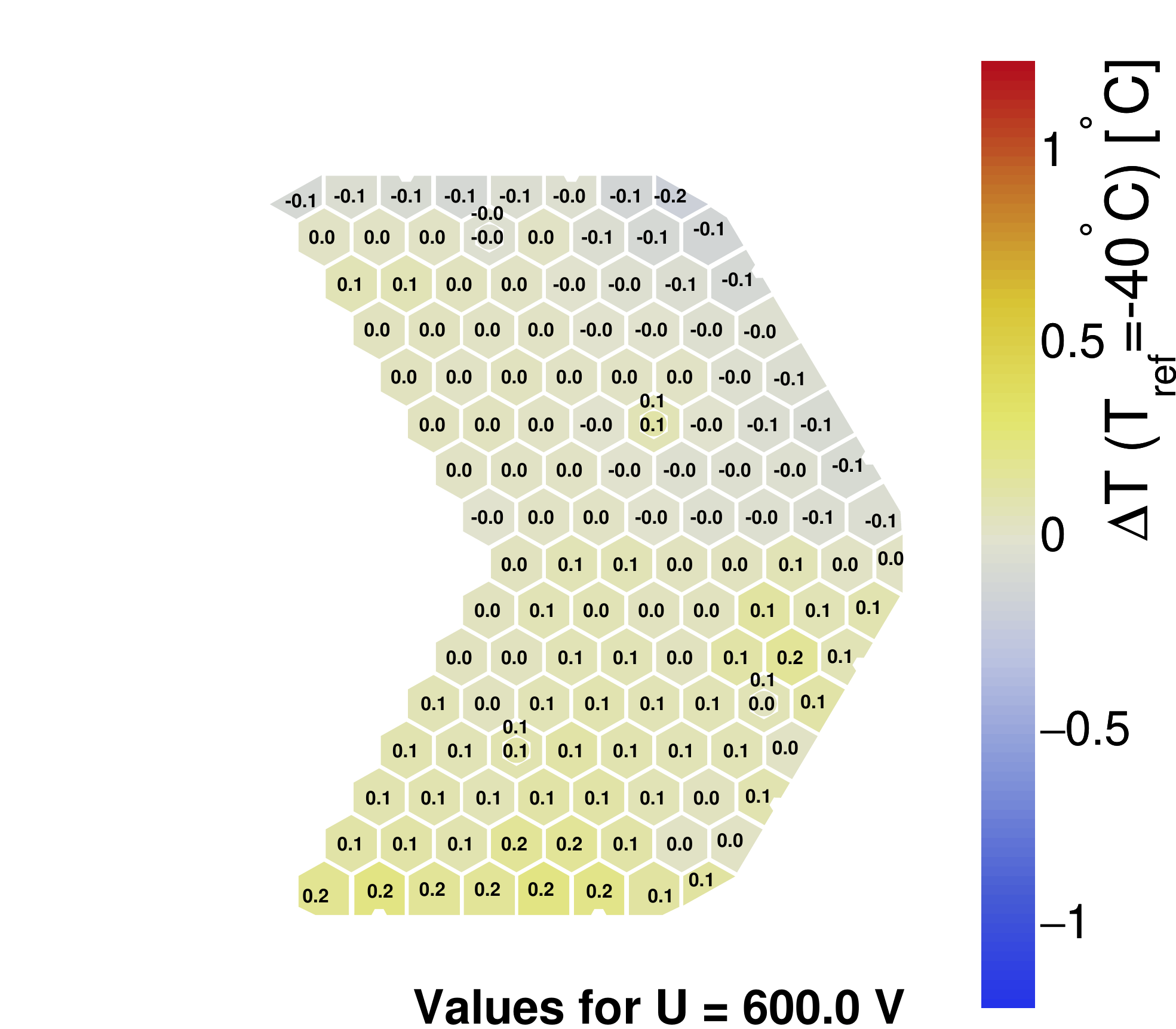}
		\subcaption{
		}
		\label{plot:chucktemp_after}
	\end{subfigure}
	\caption{
		(a) Temperature differences derived from per-pad leakage currents between symmetric locations of a representative neutron-irradiated CE silicon sensors on the cold chuck at CERN.
		(b) Determined profile of temperature differences, modelled as a two-dimensional gaussian, across the sensor surface.
		(c) Closure check: derived temperature differences after accordingly correcting the measured per-pad leakage currents.
	}
\end{figure}

The leakage current in silicon devices dependends on the temperature.
For the investigated sensors in this work, the current is mainly generated in the bulk for which the temperature ($T$) and current ($I$) are related by~\ref{eq:temp_scaling}~\cite{Chilingarov_2013}, where $E_\text{eff}=\SI{1.21}{\electronvolt}$ is the effective gap energy and $k_b$ denotes the Boltzmann constant.
\begin{equation}
    \frac{I}{T^2}\cdot \exp{\frac{E_\text{eff}}{2\cdot k_b \cdot T}}~\equiv~\text{const.}
    \label{eq:temp_scaling}
\end{equation}
The validity of this temperature scaling law applied for the CE silicon pad sensors in the relevant temperature range has been verified and is illustrated in~\ref{plot:iv_tempscaling}.
Notably, a change by \SI{1}{\celsius} impacts the current by more than 10$~\%$. 
Variations of the temperature across the actively cooled-down chuck, and with it across the sensor, may mimic neutron fluence profiles as in~\ref{plot:iv_hexplot}. 
Thus, they should be identified and corrected for. 
In this work, the temperature non-uniformity of the cold chuck at CERN (C200-40 model produced by Systems att) could be estimated from leakage current data of neutron-irradiated sensors.
By comparing per-pad currents ($I_{i(j,k)}$) at a fixed bias voltage between the symmetry locations $i$, $j$, $k$, described below, temperature differences $\delta I_{i(j), k}$ can be calculated according to~\ref{eq:temp_diff}.
For this purpose, one of the neutron-irradiated low-density sensors was characterised three times at \SI{0}{\degree}, \SI{120}{\degree}, and \SI{240}{\degree} rotation.
The precision in repeating the sensor placement on the chuck can be neglected with respect to the size of the pads. 
\begin{equation}
    \frac{\delta I_{i(j),k}}{I_k} = \frac{\delta T_{i(j), k}}{T_k} \cdot \left(2 + \frac{E_\text{eff}}{2\cdot k_b \cdot T_k} \right)~,~~T_k \coloneqq T_\text{ref}
    \label{eq:temp_diff}
\end{equation}
\ref{plot:chucktemp_before} shows the hereby computed chuck temperature differences.
The determined variation is consistent with the $\pm\SI{0.5}{\celsius}$ non-uniformity as specified by the chuck producer.
Assuming a reference temperature of $T_\text{ref}\equiv\SI{-40}{\celsius}$, a two-dimensional gaussian parameterisation bound to $[\SI{-0.5}{\celsius}, \SI{0.5}{\celsius}]$ is fitted to reproduce the temperature differences $\delta I_{i(j),k}$.
The result of the fit is shown in~\ref{plot:chucktemp_correction}.
The hereby constructed map of chuck temperature differences is input to correct the measured per-pad leakage currents using~\ref{eq:temp_scaling}.
As a closure check, the re-application of~\ref{eq:temp_diff} on the temperature-corrected dataset demonstrates the applicability of this particular chuck temperature model as shown in~\ref{plot:chucktemp_after}.